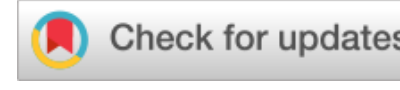

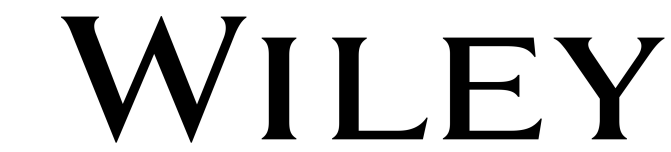

*Research Article*

# Utilizing Entropy to Systematically Quantify the Resting-Condition Baroreflex Regulation Function

**Bo-Yuan Li**,[1,2] **Xiao-Yang Li**,[1,2] **Xia Lu**,[3] **Rui Kang**,[1,2] **Zhao-Xing Tian**,[4] **and Feng Ling**[3]

[1]*School of Reliability and Systems Engineering, Beihang University, Beijing 100191, China*
[2]*Science and Technology on Reliability and Environmental Engineering Laboratory, Beijing 100191, China*
[3]*Department of Neurosurgery, Xuanwu Hospital,*
*Capital Medical University and China International Institution of Neuroscience, Beijing 100053, China*
[4]*Department of Emergency, Beijing Jishuitan Hospital, Beijing 100035, China*

Correspondence should be addressed to Xiao-Yang Li; leexy@buaa.edu.cn and Xia Lu; luxia@xwhosp.org

Baroreflex is critical to maintain blood pressure homeostasis, and the quantification of baroreflex regulation function (BRF) can provide guidance for disease diagnosis, treatment, and healthcare. Current quantification of BRF such as baroreflex sensitivity cannot represent BRF systematically. From the perspective of complex systems, we regard that BRF is the emergence result of fluctuate states and interactions in physiological mechanisms. Therefore, the three-layer emergence is studied in this work, which is from physiological mechanisms to physiological indexes and then to BRF. On this basis, since the entropy in statistical physics macroscopically measures the fluctuations of system's states, in this work, the principle of maximum entropy is adopted, and a new index called PhysioEnt is proposed to quantify the fluctuations of four physiological indexes, i.e., baroreflex sensitivity, heart rate, heart rate variability, and systolic blood pressure, which aims to represent BRF in the resting condition. Further, two datasets with different subjects are analyzed, and some new findings can be obtained, such as the contributions of the physiological interactions among organs/tissues. With measurable indexes, the proposed method is expected to support individualized medicine.

## 1. Introduction

The arterial baroreceptor reflex (referred to baroreflex) is one of mechanisms for maintaining blood pressure (BP) homeostasis, and the baroreflex regulation function, i.e., BRF, is to restore the deviated BP to optimal level in short time. The weakening or failure of BRF plays an important role in various diseases: not only chronic diseases such as hypertension, postural hypotension, and paroxysmal syncope [1], but also some acute lethal conditions such as stroke [2] and myocardial infarction [3]. If we can figure out some quantitative indexes to represent BRF and model the relationships between BRF and some measurable physiological indexes related to baroreflex, we can provide support for individual diagnosis, treatment, and health management of related diseases.

To quantify BRF, some studies focused on typical physiological indexes and construct relation curves between these indexes based on experiments. Such relation curves include the curves of baroreflex sensitivity (BRS) with peripheral osmolality [4] and blood volume [5], and the curves of BP with carotid sinus pressure [6], R-R interval (RRI) [7] and nerve activity [8, 9]. However, these physiological indexes cannot systematically characterize BRF. For instance, a typical quantification of BRF, RBS, has been proved that it mainly reflect the function of efferent nerves but not afferent pathway [10, 11]. In addition, the relation curves obtained by experiments cannot clearly correspond to specific physiological mechanisms, which limit the appliances in disease diagnosis and treatment.

Some studies adopted differential equations and automatic control models to quantify BRF based on physiological mechanisms, such as the modeling of cardiovascular and autonomic nervous system [12–16], and coupling modeling with other physiological systems [17–19]. However, such researches still have some problems: due to the complexity of human physiological mechanisms, the accurate modeling requires a large amount of detailed data and information which are difficult to obtain limited by experimental conditions. Specifically, when determining the parameters of a control element, the measurements in an open-loop case are necessary to exclude the influences of other control elements. For this purpose, the normal regulation would be blocked, and certain organs would be isolated out [20]. However, such experiments are forbidden for human ethically. Many researches applied the results from animal experiments instead [21, 22], but the applicability of these results to human is quite doubtful.

Some studies utilized information-based methods [23–26] to analyze information flows and causal relationships among physiological indexes, including BP, pulse interval, and sympathetic nervous activity. In addition, the concept of physiological network was proposed to model the relationships among indexes [27]. These studies constructed the interactions related to measurable indexes, which provided concise descriptions of complex physiological mechanisms. However, these researches still lacked a systematic representation of BRF over interactions.

From the above literature review, it can be realized that we still lack a quantitative relationship between a systematic index of BRF and measurable physiological indexes. Actually, some medical studies have identified the relationships between the fluctuations of measurable physiological indexes and the resting-condition baroreflex. Spontaneous fluctuations of the indexes related to the resting-condition baroreflex without external perturbations have been reported [28, 29], which are considered to originate from intrinsic regulation mechanisms [30, 31]. Further, the fluctuations of different indexes may indicate different physiological mechanisms and health conditions. For instance, the fluctuation of BP is considered as a signal of regulation impairment [32] and organ damage [33]. In contrast, the fluctuations of heart rate (HR) and neural activity generally refer to the responsiveness to environmental stimuli [32, 34, 35], and many pathological conditions are accompanied by an alteration or disappearance of such fluctuations.

The above physiological findings and statistical physics inspire our thinking: we can systematically quantify BRF directly based on physiological mechanisms and easily measured data of physiological indexes in clinic and daily life by using the concept of entropy. In statistical physics, entropy is a systematical characterization of microscopic fluctuation and diversity, and it has been successfully adopted to describe the functions of various complex systems [36–38]. Therefore, entropy can be adopted to construct the connection between BRF and physiological mechanisms.

In this work, we focus on the resting-condition BRF and aim to study the emergence from the baroreflex mechanisms to physiological indexes and then to BRF. Specifically, in this work, physiological entropy (PhysioEnt) is proposed to quantify the fluctuations of four indexes, i.e., systolic blood pressure (SBP), HR, heart rate variability (HRV), and BRS. To calculate PhysioEnts of indexes, the principle of maximum entropy (MaxEnt) is adopted to construct a MaxEnt model.

As a general logical foundation, the principle of MaxEnt infers the knowledge of complex systems based on given information, and provides rational and conservative solutions for systems' distributions [39, 40]. Nowadays, MaxEnt modeling has been widely adopted to study the biosystems with regulation and analyze the homeostasis with fluctuations of systems' status. For instance, it has been adopted to study the flight of bird flocks [41], in which the direction of each bird is adjusted dynamically under the interactions with other birds and the whole flock shows an orderly and steady movement. In addition, MaxEnt is also applied in the analysis on gene regulatory, such as the Saccharomyces exhibiting metabolic oscillations due to dynamic gene interactions [42]. For physiology regulatory systems, it has been used to model the couplings in RRI related to time delay, which finds the long-term correlation for couplings and may support cardiac failure classification [43]. Further, MaxEnt also has been applied to study the interactions between neurophysiological activities and evaluate the homeostasis based on energy landscape analysis [44, 45]. Conclusively, MaxEnt can be rationally utilized to model baroreflex and explore physiological findings.

In this work, two open-source datasets, Eurobavar dataset and the dataset from Jena University Hospital (Jena dataset for short), are adopted, aiming to study the BRF in different subjects and verify the proposed method. The medical findings are expected to support the diagnosis, treatment, and healthcare of related diseases. Furthermore, since the proposed methods in this work are based on measurable physiological indexes, they can be adopted briefly in daily life and support individualized healthcare.

## 2. Emergence in the Resting-Condition BRF

Baroreflex is a typical feedback regulation, and the physiological mechanisms of baroreflex can be concluded as: when BP fluctuates, blood vessels stretch, and baroreceptors are stimulated. Nerve impulses are integrated in nerve centers and affect neural activities. Then sinus nodes and other effectors are innervated to achieve a short-term regulation of BP. In the resting condition, studies have shown that sympathetic activity is low and stable, and autonomic activity is dominated by parasympathetic nerves [46, 47]. Also, studies have found that there is a lack of relationship between sympathetic activity and BP in rest [48, 49], of which the mechanisms are still unclear. From another side, for current studies evaluating sympathetic activity, a basic assumption is generally preset that sympathetic activity is mainly affected by external stimulus [50, 51]. By contrast, even in rest, parasympathetic activity can be influenced by spontaneous breathing [52]. Therefore, in this work, we mainly consider the effects of parasympathetic activity.

To represent the mechanisms of baroreflex, some physiological indexes are proposed. Physiological indexes represent whether the organs/tissues in baroreflex achieve their functions, and also reflect whether the physiological processes among organs/tissues are normal. Among various indexes, BRS is a typical index to measure the behavior of baroreflex. However, essentially, BRS describes the reflex effect of sinus nodes in response to nerve impulses, but cannot accurately reflect how nerve impulses generate [53]. In other words, although BRS involves comprehensive autonomic neural function, it mainly reflects the efferent pathway that sinus nodes are innervated, but cannot accurately represent the process in afferent pathway [10]. For this point, HRV is generally adopted to describe nerve impulses and autonomic function not only for baroreflex, thus it can reflect the process that baroreceptors stimulate parasympathetic nerves in afferent pathway [54]. In addition to the functions of nerves in baroreflex, the functions of effectors, i.e., sinus nodes and blood vessels should be considered. The function of sinus nodes is to response to autonomic nervous impulses and cause heartbeats. Therefore, HR can be adopted to reflect the combination of intrinsic sinus node function and autonomic nerves impulses. As the regulated objective, SBP reflects the resting-condition homeostasis achieved by baroreflex. Also, SBP can characterize hemodynamics and vascular mechanics including vascular elasticity, the stretching of vessel walls, and vascular resistance [55, 56].

According to the above analysis, four measurable physiological indexes, BRS, HRV, HR, and SBP can cover the main organs/tissues and physiological processes in the resting-condition baroreflex. Neither of them, including BRS, can comprehensively represent the regulation function alone, but the synergy and interaction among these indexes can emerge the resting-condition BRF. Therefore, there is an emergence from physiological mechanisms to physiological indexes and then to BRF. Specifically, baroreflex physiological mechanisms are in the bottom layer, which contain specific organs/tissues and physiological processes. Physiological indexes belong to the middle layer, and the interactions among indexes should be considered due to the coupling of mechanisms. Finally, the resting-condition BRF is in the top layer. Such an emergence process is conceptualized in Figure 1, which is expected to be described and modeled in this work.

According to the review in Section 1, it is reasonable to regard that the fluctuations of BRS, SBP, HR, and HRV characterize the resting-condition BRF. If we can quantify the fluctuations of four indexes, we can quantitatively evaluate the resting-condition BRF based on these indexes.

## 3. Methods

*3.1. Physiological Entropy and Maximum Entropy Model.* Entropy is the foundation of statistical physics and measures the diversity of a thermodynamic system. Shannon developed the entropy in information theory, which measures the diversity inherent to the states of systems. The entropy of a discrete random variable $X$ was proposed first, and then the concept of entropy was extended by Shannon to continuous condition, as shown in the following equation:

$$H(X) = \begin{cases} -\sum_{i=1}^{n} P(x_i) \ln P(x_i), & \text{if } X \text{ is discrete} - \text{valued}, \\ -\int P(x) \ln P(x), & \text{if } X \text{ is continuous} - \text{valued}, \end{cases} \tag{1}$$

where $H(X)$ is the entropy of $X$ and $P$ is the probability of $X$.

Further, Jaynes figured out that statistical physics is only to do probabilistic inferences from limited information and combined the entropy in statistical physics and information theory [39, 40]. In a word, entropy macroscopically measures how a system's microscopic diversity emerges system's behaviors and the capacities of system's functions. Therefore, it can quantify the fluctuating states of four physiological indexes; consequently, construct the connection between BRF and physiological mechanisms.

The states of four physiological indexes, i.e., BRS, SBP, HR, and HRV, are continuous-valued variables. However, when calculating the entropy of continuous-valued variables, the continuous entropy in equation (1) has been proved to be problematic [57]. Therefore, in this work, continuous indexes are normalized and partitioned into fixed value spaces for discretization, which is further introduced in section 3.3. Through data preprocessing, discrete indexes can be obtained, and for each index, the physiological entropy (PhysioEnt for short) can be defined as

$$S^{\gamma} = -\sum_{x_{\gamma}} P(x_{\gamma}) \ln P(x_{\gamma}), \gamma = \text{SBP, HR, HRV, BRS}, \tag{2}$$

where $P(x_{\gamma})$ is the marginal probability distribution of index $x_{\gamma}$.

From equation (2), it can be figured that the more diverse state of one index is, the more significant its fluctuation is; and then the larger its PhysioEnt is; otherwise, the smaller PhysioEnt corresponds to the less significant fluctuation. According to current studies, an increase in the fluctuation of SBP generally represents a compromised internal homeostasis. Meanwhile, the fluctuation of HR reflects the responsiveness of effectors including sinus nodes to various stimulus. And the fluctuations of HRV and BRS can represent the responsiveness of parasympathetic nerves to stimulus, respectively. Therefore, it can be further concluded that a worse health condition of the resting-condition BRF may correspond to a higher PhysioEnt of SBP and lower PhysioEnts of HR, HRV, and BRS.

According to equation (2), the estimation of $P(x_{\gamma})$ is necessary to calculate PhysioEnts. However, for the complex baroreflex system, its inherent mechanisms are still difficult to precisely understood and modeled, and it is quite challenging to perform a reductionist, bottom-up modeling for $P(x_{\gamma})$. For this problem, the principle of MaxEnt provides a general logical foundation from another aspect. That is, start from the phenomena and behaviors of systems, understand the internal laws of systems from top to bottom, and then predict systems' status [58]. MaxEnt modeling uses the possessed information and knowledge of systems as constraints, and estimates distributions by maximizing Shannon information entropy, which aims to eliminate any subjective bias for distribution assumptions. Therefore, MaxEnt modeling provides brief and conservative solutions for systems and concludes new knowledge from statistical data.

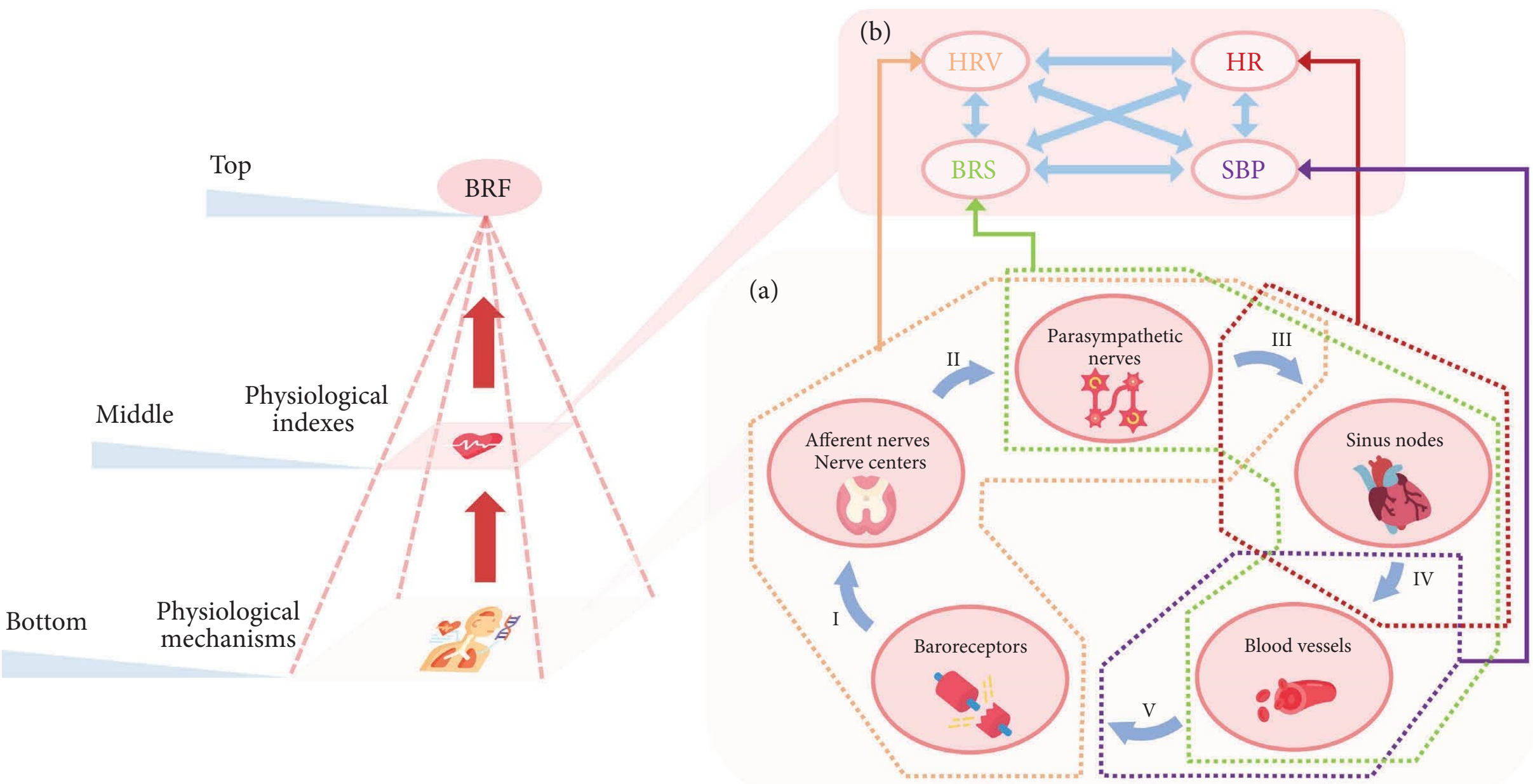

FIGURE 1: Emergence from physiological mechanisms to physiological indexes and BRF. (a) For physiological mechanisms, nodes indicate organs/tissues, and arrows indicate physiological processes, where (I) impulses generated by baroreceptors transmit through the afferent pathway and integrate in nerve centers; (II) nerve centers innervate parasympathetic activities; (III) parasympathetic nerves innervate sinus nodes to change HR; (IV) HR influences BP and the stretching of blood vessels; (V) baroreceptors sense the stretching. Dashed parts indicate the emerged indexes corresponding to different organs/tissues and processes. (b) Physiological indexes, BRS, SBP, HR, and HRV, emerge from physiological mechanisms. Arrows indicate the possible interactions among indexes.

It should be figured that the modeling process of MaxEnt is essentially eliminating unreasonable subjective biases when we understand systems, rather than constructing inherent mechanisms analytically. In other words, the optimized entropy during modeling indicates the uncertainty reasonably retained from the aspect of information when we infer knowledge from data, but not the uncertainty or diversity generated from real mechanisms of system. Therefore, MaxEnt modeling does not rely on specific mechanisms of systems to deal with uncertainty, such as maintaining stability or tending to divergence. Rather, some systems' mechanisms are exactly the knowledge inferred from data by MaxEnt modeling.

In this work, the principle of MaxEnt is adopted to estimate $P(x_\gamma)$ by inferring the joint probability distribution of $x = \{x_{\mathrm{BRS}}, x_{\mathrm{SBP}}, x_{\mathrm{HR}}, x_{\mathrm{HRV}}\}$, i.e., $P(x)$. In a MaxEnt model, the interactions among factors are represented by the model components composed of these factors. In this work, to calculate PhysioEnts and reflect physiological mechanisms, the estimated $P(x)$ based on a MaxEnt model, i.e., $\widehat{P}(x)$, can be written as follows:

$$\widehat{P}(x) = \frac{1}{Z}\exp\left[\sum_{\mu=1}^{N} \lambda^{\mu} f^{\mu}(x)\right], \tag{3}$$

where $Z$ is a parameter for normalization, $f^{\mu}(x)$ means the $\mu^{\mathrm{th}}$ model component, $\lambda^{\mu}$ is the model parameter corresponding to $f^{\mu}(x)$, and $N$ is the total number of components.

Different model components represent different interdependencies among indexes, and will lead to different MaxEnt models. Besides the independent effects reflected by $x_{\mathrm{BRS}}$, $x_{\mathrm{SBP}}$, $x_{\mathrm{HR}}$, and $x_{\mathrm{HRV}}$, the interactions between two indexes can be represented by the model components obtained by multiplying two indexes, i.e., $x_{\mathrm{BRS}}x_{\mathrm{SBP}}$. Similarly, the interactions among three indexes and four indexes can also be represented by model components such as $x_{\mathrm{BRS}}x_{\mathrm{SBP}}x_{\mathrm{HR}}$ and $x_{\mathrm{BRS}}x_{\mathrm{SBP}}x_{\mathrm{HR}}x_{\mathrm{HRV}}$. The models considering interactions between two indexes, among three indexes and among four indexes are respectively called Pairwise model $\widehat{P}_{\mathrm{II}}(x)$, Triplet model $\widehat{P}_{\mathrm{III}}(x)$ and Quaternion model $\widehat{P}_{\mathrm{IV}}(x)$. For example, the specific form of $\widehat{P}_{\mathrm{II}}(x)$ is

$$
\begin{aligned}
\widehat{P}_{\mathrm{II}}(x) &= \frac{1}{Z_{\mathrm{II}}}\exp\left[\sum_{\mu=1}^{10}\lambda_{\mathrm{II}}^{\mu} f_{\mathrm{II}}^{\mu}(x)\right] \\
&= \frac{1}{Z_{\mathrm{II}}}\exp\left[\lambda_{\mathrm{II}}^{1} x_{\mathrm{BRS}} x_{\mathrm{SBP}} + \lambda_{\mathrm{II}}^{2} x_{\mathrm{BRS}} x_{\mathrm{HR}} + \lambda_{\mathrm{II}}^{3} x_{\mathrm{BRS}} x_{\mathrm{HRV}} + \lambda_{\mathrm{II}}^{4} x_{\mathrm{SBP}} x_{\mathrm{HR}} + \lambda_{\mathrm{II}}^{5} x_{\mathrm{SBP}} x_{\mathrm{HRV}}\right. \\
&\quad \left. + \lambda_{\mathrm{II}}^{6} x_{\mathrm{HR}} x_{\mathrm{HRV}} + \lambda_{\mathrm{II}}^{7} x_{\mathrm{BRS}} + \lambda_{\mathrm{II}}^{8} x_{\mathrm{SBP}} + \lambda_{\mathrm{II}}^{9} x_{\mathrm{HR}} + \lambda_{\mathrm{II}}^{10} x_{\mathrm{HRV}}\right],\ Z_{\mathrm{II}} = \sum_{\mathbf{x}} \exp\left[\sum_{\mu=1}^{10}\lambda_{\mathrm{II}}^{\mu} f_{\mathrm{II}}^{\mu}(x)\right].
\end{aligned} \tag{4}
$$

For the solutions of model parameters, $\lambda_k^{\mu}$, $k = \mathrm{II}, \mathrm{III}, \mathrm{IV}$, iterative scaling algorithm can be adopted [59]:

$$
\left(\lambda_k^{\mu}\right)_{t+1} = \left(\lambda_k^{\mu}\right)_t + \alpha \times \operatorname{sign}\left(\left\langle f_k^{\mu}(x)\right\rangle_D\right) \times \ln\left(\frac{\left\langle f_k^{\mu}(x)\right\rangle_D}{\left\langle f_k^{\mu}(x)\right\rangle_t}\right), \tag{5}
$$

where $(\lambda_k^{\mu})_t$ is the $\lambda_k^{\mu}$ after $t^{\mathrm{th}}$ iteration; $\langle f_k^{\mu}(x)\rangle_D$ is the mean of $f_k^{\mu}(x)$ calculated by data, $\langle f_k^{\mu}(x)\rangle_t$ is the mean of $f_k^{\mu}(x)$ in the model after $t^{\mathrm{th}}$ iteration, $\alpha$ is the learning rate that is set as 0.75 and $\operatorname{sign}(\bullet)$ is signum function which is defined as $\operatorname{sign}(x) = \begin{cases} -1 \text{ if } x < 0, \\ 0 \text{ if } x = 0, \\ 1 \text{ if } x > 0. \end{cases}$ The stop condition of iteration is that: $\sum_{\mu=1}^{N_k}\left\{ (|\langle f_k^{\mu}(x)\rangle_D - \langle f_k^{\mu}(x)\rangle_t| / \langle f_k^{\mu}(x)\rangle_D) \times 100\% \right\} \leq 0.5\%$, where $N_k$ is the total number of the model components. $\langle f_k^{\mu}(x)\rangle_t$ is calculated by Metropolis–Hastings algorithm [60].

Based on $\widehat{P}(x)$, the marginal distribution of each index can be calculated. Take $\widehat{P}_{\mathrm{II}}(x)$ for instance, and the marginal distribution of one discrete-valued index $x_{\gamma}$ can be calculated as follows:

$$
\widehat{P}_{\mathrm{II}}\left(x_{\gamma}\right) = \sum_{\mathbf{x}/x_{\gamma}} \widehat{P}_{\mathrm{II}}(x), \tag{6}
$$

where $x/x_{\gamma}$ represents the vector of indexes except $x_{\gamma}$. Substitute equations (6) into (2), and the PhysioEnt of each physiological index can be obtained.

The rationality of MaxEnt models can be quantified based on multi-information ratio to support model selection [61]. Take $\widehat{P}_{\mathrm{II}}(x)$ for instance, its multi-information ratio $\eta_{\mathrm{II}}$ can be calculated as follows:

$$
\eta_{\mathrm{II}} = \frac{S_{\mathrm{ind}} - S_{\mathrm{II}}}{S_{\mathrm{ind}} - S_D}, \tag{7}
$$

where $S_{\mathrm{ind}}$ is the entropy of $P_{\mathrm{ind}}(x)$. $P_{\mathrm{ind}}(x)$ is the distribution considering no interaction, that is, $P_{\mathrm{ind}}(x) = P(x_{\mathrm{BRS}}) \times P(x_{\mathrm{SBP}}) \times P(x_{\mathrm{HR}}) \times P(x_{\mathrm{HRV}})$; $S_{\mathrm{II}}$ is the entropy of $\widehat{P}_{\mathrm{II}}(x)$; and $S_D$ is the entropy of observed data. $\eta_{\mathrm{II}}$ measures the similarity between $\widehat{P}_{\mathrm{II}}(x)$ and data in terms of information gain. The closer to 1 $\eta_{\mathrm{II}}$ is, the more accurate $\widehat{P}_{\mathrm{II}}(x)$ is.

For the models introducing high-order interactions, i.e., $\widehat{P}_{\mathrm{III}}(x)$ and $\widehat{P}_{\mathrm{IV}}(x)$, they would theoretically achieve larger multi-information ratios, unless there is no high-order interaction at all. However, even if there is no high-order interaction, when modeling with limited data, overfitting may still occur, which leads to an increase in multi-information ratios. Therefore, the marginal effect of the increase in multi-information ratios should be considered when introducing high-order interactions, and make a balance between the accuracy and simplicity of modeling.

### 3.2. Relative Contributions of Model Components

*3.2. Relative Contributions of Model Components.* From the perspective of model itself, it is composed of various model components. Take $\widehat{P}_{\mathrm{II}}(x)$ as an example: the model components in $\widehat{P}_{\mathrm{II}}(x)$ can be divided into interactive components including $x_{\mathrm{BRS}}x_{\mathrm{SBP}}$, $x_{\mathrm{BRS}}x_{\mathrm{HR}}$, $x_{\mathrm{BRS}}x_{\mathrm{HRV}}$, $x_{\mathrm{SBP}}x_{\mathrm{HR}}$, $x_{\mathrm{SBP}}x_{\mathrm{HRV}}$, and $x_{\mathrm{HR}}x_{\mathrm{HRV}}$, and independent components including $x_{\mathrm{BRS}}$, $x_{\mathrm{SBP}}$, $x_{\mathrm{HR}}$, and $x_{\mathrm{HRV}}$.

From Figure 1, we can figure out that these independent components correspond to the elements in physiological index layer. And these interactive components correspond to the interactions between every two elements in this layer. Further, we can regard interactive components as specific physiological processes in physiological mechanism layer (see Supplementary Texts 1 for details), and regard independent components as specific organs/tissues (see section 2 for details). Therefore, if we can figure out how much each component contributes to $\widehat{P}_{\mathrm{II}}(x)$, corresponding results can indicate how and how much the associated physiological processes or organs/tissues make effects on the resting-condition BRF.

To quantify the contribution of each model component, we propose the concept of relative contribution and its quantitative index *RC*. First, we define the concept of generalized energy, $E(x)$, as shown in equation (8), where $f_{\mathrm{II}}^{\mu}(x)$ represents the $\mu^{\mathrm{th}}$ model component and $\lambda_{\mathrm{II}}^{\mu}$ is the corresponding parameter. According to equation (4), $E(x)$ is the numerator of $\widehat{P}_{\mathrm{II}}(x)$, and the total number of model components of $\widehat{P}_{\mathrm{II}}(x)$ is 10.

$$
E(x) = \exp\left[\sum_{\mu=1}^{10}\lambda_{\mathrm{II}}^{\mu} f_{\mathrm{II}}^{\mu}(x)\right] = \prod_{\mu=1}^{10}\exp\left[\lambda_{\mathrm{II}}^{\mu} f_{\mathrm{II}}^{\mu}(x)\right]. \tag{8}
$$

Since the denominator of $\widehat{P}_{\mathrm{II}}(x)$ is a parameter for normalization, the value of $E(x)$ determines the value of $\widehat{P}_{\mathrm{II}}(x)$, and then determines the value of PhysioEnts according to equation (4). The greater the $E(x)$ is, the greater the $\widehat{P}_{\mathrm{II}}(x)$ is, and the greater possibility that the values of indexes are $x$ is.

Then, the energy proportion of the $\nu^{\text{th}}$ model component given $x$, $R_\nu(x)$, is defined. $R_\nu(x)$ represents the portion of $E(x)$ occupied by the $\nu^{\text{th}}$ model component given $x$. Since $E(x)$ is calculated by multiplying the exponential functions of all model components, $R_\nu(x)$ can be expressed as follows:

$$R_\nu(x) = \frac{\exp\left[\lambda_{\mathrm{II}}^{\nu} f_{\mathrm{II}}^{\nu}(x)\right]}{E(x)} = \frac{1}{\prod_{\mu\neq\nu} \exp\left[\lambda_{\mathrm{II}}^{\mu} f_{\mathrm{II}}^{\mu}(x)\right]}. \tag{9}$$

The greater the $R_\nu(x)$ is, the more significantly $E(x)$ depends on the $\nu^{\text{th}}$ model component. Therefore, it can be inferred that the $\nu^{\text{th}}$ model component contributes more significantly to $\widehat{P}_{\mathrm{II}}(x)$. Since PhysioEnts are determined by $\widehat{P}_{\mathrm{II}}(x)$, $R_\nu(x)$ also indicates the contribution of the $\nu^{\text{th}}$ model component to PhysioEnts.

Based on $R_\nu(x)$, $RC_\nu$ is defined as the result after normalizing for all model components and averaging for all $x$, as shown in equation (10). Essentially, $RC_\nu$ reflects the contribution of the $\nu^{\text{th}}$ model component to $\widehat{P}_{\mathrm{II}}(x)$ and PhysioEnts.

$$RC_\nu = \sum_x \frac{R_\nu(x)}{\sum_{\mu=1}^{10} R_\mu(x)} \widehat{P}_{\mathrm{II}}(x). \tag{10}$$

For interactive components, a large $RC$ indicates that the interaction between the associated two indexes contributes more to BRF. Furthermore, this may suggest BRF depends more on corresponding physiological process. For independent components, a larger $RC$ indicates that corresponding index makes a more significant independent effect on BRF. And this may suggest that BRF depends more on the function of corresponding organs/tissues. In other words, corresponding organs/tissues may be less susceptible to other organs/tissues.

*3.3. Data Description and Preprocessing.* In this work, Eurobavar dataset and Jena dataset are analyzed to study the BRF in different subjects and verify the proposed method.

*3.3.1. Eurobavar Dataset.* Eurobavar dataset [62] is a widely used dataset to study baroreflex regulation and test new methods for BRF evaluation, which includes 21 subjects. The subjects contain 17 females and 4 males, and the mean age is 38.4 with a standard error of 3.3 years. Among these subjects, two subjects, B005 and B010, are baroreflex-impaired, due to a diabetic autonomic neuropathy and a heart transplant procedure, respectively. And this can support the analysis in the context of baroreflex dysfunction.

Eurobavar dataset contains the recordings of interbeat intervals (IBI) derived from ECG recordings measured with a Datex cardiocap II monitor; and continuous noninvasive beat-to-beat SBP series, which was recorded using a Finapres 2300 device. The subjects were recorded in supine position and standing position. The data were recorded in a quiet room, at constant temperature (20°C) and luminosity, and any disturbance such as noise or entry in the room was avoided. For the resting-condition BRF, the data in supine position is analyzed in this work.

*3.3.2. Jena Dataset.* Jena dataset contains the resting recordings of ECG and continuous noninvasive BP of 1121 healthy subjects [63, 64]. The criterion of healthy subjects is that the subjects do not have any medical conditions, illegal drugs, or medication potentially influencing cardiovascular function. And there is no pathological finding according to thorough physical examination, resting electrocardiography and routine laboratory parameters (electrolytes, basic metabolic panel, and blood count).

In Jena dataset, ECG recordings were measured by a MP150 or Task Force Monitor system, and the noninvasive BP was recorded by a Task Force Monitor system or CNAP 500. The sampling frequency of each recording is 1000 Hz. Measurements were performed in a room with the controlled temperature 22°C and constant illumination level, and subjects were in supine position. Conclusively, the strict examination and record procedure ensure that the data can reflect the spontaneous resting-condition regulation of healthy subjects.

The labels of Jena subjects include age, sex, and BMI. The ages range between 18 and 92 are divided into 15 groups to ensure that subjects cannot be identified based on demographic information. Age groups are shown in Table 1.

*3.3.3. Data Preprocessing.* The recordings of both datasets are first preprocessed by following methods to obtain HR, SBP, HRV, and BRS.

(1) HR: For Eurobavar dataset, HR can be calculated by $X_{\mathrm{HR}}^{i}(j) = (60/X_{\mathrm{IBI}}^{i}(j))$, where $X_{\mathrm{IBI}}^{i}(j)$ represents the $j^{\text{th}}$ $X_{\mathrm{IBI}}^{i}(j)$ IBI of the $i^{\text{th}}$ subject. And the time of $X_{\mathrm{IBI}}^{i}(j)$, $t^i(j)$, can be obtained by $t^i(j) = \sum_{u\leq j-1} X_{\mathrm{IBI}}^{i}(u) + X_{\mathrm{IBI}}^{i}(j)/2$. For Jena dataset, the RRIs of each subject are first obtained from ECG utilizing the mhrv toolbox in MATLAB [65], and $X_{\mathrm{R\text{-}R}}^{i}(j)$ is the $j^{\text{th}}$ $X_{\mathrm{R\text{-}R}}^{i}(j)$ calculated RRI of the $i^{\text{th}}$ subject. Corresponding time series can also be obtained from the mhrv toolbox: $t^i = \{t^i(j)\}$. Then, HR can be calculated by $X_{\mathrm{HR}}^{i}(j) = (60/X_{\mathrm{R\text{-}R}}^{i}(j))$. The calculated HR can be expressed as $X_{\mathrm{HRHR}}^{i} = \{X_{\mathrm{HR}}^{i}(j)\}$.
(2) SBP: For Eurobavar dataset, SBP has been recorded. For Jena dataset, the maximum value of the BP signal between $t^i(j)$ and $t^i(j+1)$ is taken as $X_{\mathrm{SBP}}^{i}(j)$. Consequently, SBP data can be referred as $X_{\mathrm{SBP}}^{i} = \{X_{\mathrm{SBP}}^{i}(j)\}$.
(3) HRV: For these two datasets, HRV can be represented indirectly by the standard deviation of RRI or IBI [66]. Therefore, a time window $T_{\mathrm{HRV}}$ is set and HRV can be calculated in $T_{\mathrm{HRV}}$. First, for $t^i(j)$, we check the time points before and after $t^i(j)$, and find the time points $t^i(j-p)$ and $t^i(j+q)$ satisfying $t^i(j+q)-t^i(j)\geq (T_{\mathrm{HRV}}/2)$ and $t^i(j)-t^i(j-p)\geq (T_{\mathrm{HRV}}/2)$. Therefore, $t^i(j)$ can be approximate to the midpoint of time window from $t^i(j-p)$ to $t^i(j+q)$. And the standard deviation of interval (IBI in Eurobavar dataset or RRI in Jena dataset) from $t^i(j-p)$ to $t^i(j+q)$ is utilized as HRV at $t^i(j)$. As a result, the time window is sliding with time and each $t^i(j)$ corresponds to its own time window and calculated HRV. Finally, $X^i_{HRV}=\{X^i_{\mathrm{HRV}}(j)\}$ can be obtained. $T_{\mathrm{HRV}}$ is set as 10s [66].

Table 1: Age groups and corresponding ages in Jena dataset.

| Age group | Age (years) |
|---|---|
| 1 | 18~19 |
| 2 | 20~24 |
| 3 | 25~29 |
| 4 | 30~34 |
| 5 | 35~39 |
| 6 | 40~44 |
| 7 | 45~49 |
| 8 | 50~54 |
| 9 | 55~59 |
| 10 | 60~64 |
| 11 | 65~69 |
| 12 | 70~74 |
| 13 | 75~79 |
| 14 | 80~84 |
| 15 | 85~92 |

(4) BRS: For these two datasets, BRS can be calculated in frequency domain by equation (11) [67]:

$$X^i_{\mathrm{BRS}}(j)=\left(\frac{p^i_{\mathrm{interval}}(j)}{p^i_{\mathrm{SBP}}(j)}\right)^{1/2}, \tag{11}$$

where $p^i_{\mathrm{interval}}(j)$ is the spectral power of interval (IBI in Eurobavar dataset or RRI in Jena dataset) and $p^i_{\mathrm{SBP}}(j)$ is the spectral power of SBP, which are both from 0.07 Hz to 0.14 Hz. Further, a time window $T_{\mathrm{BRS}}$ also can be set to calculate BRS. First, for $t^i(j)$, $t^i(j-p)$, and $t^i(j+q)$ are identified, respectively, which satisfies $t^i(j+q)-t^i(j)\geq (T_{\mathrm{BRS}}/2)$ and $t^i(j)-t^i(j-p)\geq (T_{\mathrm{BRS}}/2)$. Then the spectral powers of interval and SBP from $t^i(j-p)$ to $t^i(j+q)$ are calculated to deduce $X^i_{\mathrm{BRS}}(j)$, and finally $X^i_{\mathrm{BRS}}=\{X^i_{\mathrm{BRS}}(j)\}$. $T_{\mathrm{BRS}}$ is set as 120s [67].

Considering the data missing at the start and end of data series due to time windows, there is a truncation so that the data series $X^i_{\mathrm{HR}}$, $X^i_{\mathrm{SBP}}$, $X^i_{\mathrm{HRV}}$, and $X^i_{\mathrm{BRS}}$ have same lengths. The examples of four time series in two datasets are given as Figures S1 and S2 in Supplementary Materials.

For the MaxEnt modeling, the normalization is conducted first according to equation (12), which aims to eliminate the influence of index dimensions:

$$\begin{aligned}
x^i_{\mathrm{BRS}}(j)&=\frac{X^i_{\mathrm{BRS}}(j)-\min\left\{X^i_{\mathrm{BRS}}\right\}}{\max\left\{X^i_{\mathrm{BRS}}\right\}-\min\left\{X^i_{\mathrm{BRS}}\right\}}\\
x^i_{\mathrm{SBP}}(j)&=\frac{X^i_{\mathrm{SBP}}(j)-\min\left\{X^i_{\mathrm{SBP}}\right\}}{\max\left\{X^i_{\mathrm{SBP}}\right\}-\min\left\{X^i_{\mathrm{SBP}}\right\}}\\
x^i_{\mathrm{HR}}(j)&=\frac{X^i_{\mathrm{HR}}(j)-\min\left\{X^i_{\mathrm{HR}}\right\}}{\max\left\{X^i_{\mathrm{HR}}\right\}-\min\left\{X^i_{\mathrm{HR}}\right\}}\\
x^i_{\mathrm{HRV}}(j)&=\frac{X^i_{\mathrm{HRV}}(j)-\min\left\{X^i_{\mathrm{HRV}}\right\}}{\max\left\{X^i_{\mathrm{HRV}}\right\}-\min\left\{X^i_{\mathrm{HRV}}\right\}},
\end{aligned} \tag{12}$$

where $x^i_{\mathrm{BRS}}(j)$, $x^i_{\mathrm{SBP}}(j)$, $x^i_{\mathrm{HR}}(j)$, $x^i_{\mathrm{HRV}}(j)$ represent the $j^{\mathrm{th}}$ normalized BRS, SBP, HR, and HRV data of the $i^{\mathrm{th}}$ subject, respectively.

To calculate PhysioEnts, the normalized data is further discretized. The value space of each index is coarse grained into 10 uniform levels from 0 to 1, and continuous-valued data points are assigned the midpoints of the levels they fall into. For instance, the data points falling into [0, 0.1), [0.1, 0.2), and [0.2, 0.3) are assigned 0.05, 0.15, and 0.25, respectively. It should be figured that the discretization would not affect the analysis on PhysioEnts, which can be referred to Supplementary Texts 2 and Table S3, S4, S5, and S6 for details.

*3.4. Statistical Analysis.* Since the large-sample Jena subjects are labelled with age, sex, and BMI, the demographic features of the resting-condition BRF can be analyzed. For this purpose, multivariable linear regression is adopted to study the trends of PhysioEnts and RCs with three labels. Age, sex, and BMI are independent variables, while PhysioEnts and RCs are dependent variables. For regression results, $F$-test is performed to verify whether the regression model is statistically significant; and $t$-test is performed to identify the significance of each label's effect on the dependent variable.

In addition, the differences among model components' RCs are also tested for comparisons. A one-way ANOVA is first performed for homogeneity test of variance. When variances of *RC*s are similar, a Scheffé test is adopted as post hoc test to compare the differences of RCs' means; for unequal variances, a two-tailed Games–Howell test is utilized for pairwise comparisons.

Since multiple statistical tests are performed on Jena dataset, Benjamini–Hochberg procedure (BH procedure) is adopted to control fault discovery rate [68]. BH procedure can be summarized as follows:

(1) For the $p$ values of all tests, i.e., $p_1,\ldots,p_m$, and $m$ is the total number of tests, they are listed in ascending order and denoted as $p_{(1)},\ldots,p_{(m)}$.

(2) For a given $\alpha$ level, which is 0.05 in this work, find the largest $k$ that $p_{(k)}\leq (k/m)\alpha$.

(3) The tests corresponding to $p_{(1)}, \ldots, p_{(k)}$ are significant.

For the following statistical results, the $p$ value representing a significant result according to BH procedure is marked with $*$. The $p$ value marked with § means that it is less than 0.05 but does not show significance according to BH procedure. All statistical analyses are performed using IBM SPSS Statistics.

## 4. Results from Eurobavar Subjects

Based on the data preprocessing procedure in section 3.3, the $\widehat{P}_{\mathrm{II}}(x)$, $\widehat{P}_{\mathrm{III}}(x)$, and $\widehat{P}_{\mathrm{IV}}(x)$ for 21 subjects are calculated separately. Averaged multi-information ratios of 21 subjects for $\widehat{P}_{\mathrm{II}}(x)$, $\widehat{P}_{\mathrm{III}}(x)$, and $\widehat{P}_{\mathrm{IV}}(x)$ are 0.8829, 0.8882, and 0.8910, respectively. It means that these models can effectively represent more than 88% of the information embedded in data, and there is no significant difference among them. Therefore, $\widehat{P}_{\mathrm{II}}(x)$ can represent observed data well enough. It might provide some evidence and support for why current physiological and medical studies focus on the pairwise correlations between physiological indexes. And the following analysis is conducted based on $\widehat{P}_{\mathrm{II}}(x)$, i.e., equation (4).

On this basis, the PhysioEnts of SBP, HR, HRV, and BRS for each subject are evaluated, and the comparison between baroreflex-impaired subjects and nonimpaired ones can be performed, as depicted in Figure 2.

According to Figure 2, baroreflex-impaired subjects possess significant lower PhysioEnts of HR, HRV, and BRS. Such results indicate that the decrease in PhysioEnts of HR, HRV, and BRS can effectively represent the weakening of the resting-condition BRF, which can support BRF evaluation.

The model components' *RC*s in $\widehat{P}_{\mathrm{II}}(x)$ for each subject are also calculated. Similarly, Figure. 3 shows the *RC*s of nonimpaired subjects and those of impaired ones respectively. Further, Figure 3(a) shows the sum of interactive components' *RC*s and the sum of independent ones, and Figure 3(b) shows the *RC* of each model component.

According to Figure 3(a), for nonimpaired subjects, the sum of interactive components' *RC*s is greater than the sum of independent ones, which indicates the values of interactive components have greater effects on $\widehat{P}_{\mathrm{II}}(x)$ and PhysioEnts compared with independent components. Further, this may suggest that for nonimpaired subjects, the interactive processes among organs/tissues affect more significantly to the resting-condition BRF, compared with independent functions of organs/tissues. However, it is exactly opposite for impaired subjects, which is mainly due to the significant great *RC* of $x_{\mathrm{HR}}$ according to Figure 3(b).

For the above results of Eurobavar subjects, detailed analysis is provided in section 6.1.

## 5. Results from Jena Subjects

For Jena dataset, the $\widehat{P}_{\mathrm{II}}(x)$, $\widehat{P}_{\mathrm{III}}(x)$, and $\widehat{P}_{\mathrm{IV}}(x)$ of 1058 subjects are calculated, in which we censor the data of 63 subjects with large estimation errors or missing labels. Averaged multi-information ratios of 1058 subjects for $\widehat{P}_{\mathrm{II}}(x)$, $\widehat{P}_{\mathrm{III}}(x)$, and $\widehat{P}_{\mathrm{IV}}(x)$ are 0.8810, 0.8822, and 0.8832, respectively, which also indicates $\widehat{P}_{\mathrm{II}}(x)$ is appropriate to describe the data. The following analysis is also conducted for $\widehat{P}_{\mathrm{II}}(x)$.

Based on $\widehat{P}_{\mathrm{II}}(x)$, PhysioEnts of indexes can be calculated. To study their demographic features, multivariable linear regressions are conducted, and detailed analysis is given in section 5.1. As for model components' *RC*s in $\widehat{P}_{\mathrm{II}}(x)$, we calculate the *RC*s of all model components for each subject. To compare the differences among *RC*s, a one-way ANOVA is first performed to test the homogeneity of variance. The $p$ value of one-way ANOVA is less than 1E-10, which indicates heterogeneity. Therefore, a two-tailed Games–Howell test is adopted to compare means of *RC*s. In addition, the demographic features of *RC*s also can be analyzed based on multivariable linear regressions. Detailed analyses are given in sections 5.2.

*5.1. Demographic Features of PhysioEnts.* Statistical regression results for PhysioEnts are depicted in Table 2, and it can be concluded that: (1) females own greater PhysioEnt of HR than male; (2) there are significant negative correlations between age and the PhysioEnts of HRV and BRS; (3) the PhysioEnt of BRS is positively correlated with BMI; and (4) although the $F$-test between age and the PhysioEnt of SBP may be not significant enough according to BH procedure, it also can indicate a relatively positive correlation between age and the PhysioEnt of SBP.

Figures 4, 5, and 6 provide the trends of PhysioEnts with age, sex, and BMI. Such demographic features are further discussed in section 6.2.1.

*5.2. Demographic Features of RCs.* For model components' *RC*s, an overview can be first performed from the aspect of interactive and independent components. The comparison between the sum of interactive components' *RC*s and the sum of independent ones can be referred as Figure 7, which indicates that the sum of interactive components' *RC*s is significantly greater than the sum of independent ones. Like the results of nonimpaired subjects in Eurobavar dataset, this may also suggest that the resting-condition BRF depends more on the physiological processes among organs/tissues.

In addition, for the comparison among various model components, the statistical results of two-tailed Games–Howell test can be referred to Table S1 and S2 in Supplementary Materials. Figure 8 provides a general comparison, and Figure 9 depicts the comparison with different age and sex groups.

Statistical results indicate that the *RC* of $x_{\mathrm{BRS}}x_{\mathrm{HRV}}$ is significantly great among interaction components. For independent components, the *RC* of $x_{\mathrm{SBP}}$ is significantly great and the *RC* of $x_{\mathrm{HRV}}$ is the smallest one. Such results are further discussed in section 6.2.1.

On the basis of Figure 9, the demographic features of *RC*s can be statistically studied. Regression results show that the sum of interactive components' *RC*s and the sum of independent ones have significant correlations with sex, as provided in Table 3 and Figure 10. The sum of interactive components' *RCs* is greater for males, while the sum of independent ones is greater in females. This indicates that the physiological processes of males contribute more to their resting-condition BRF than those of females, while the organs/tissues of females contribute more than those of males.

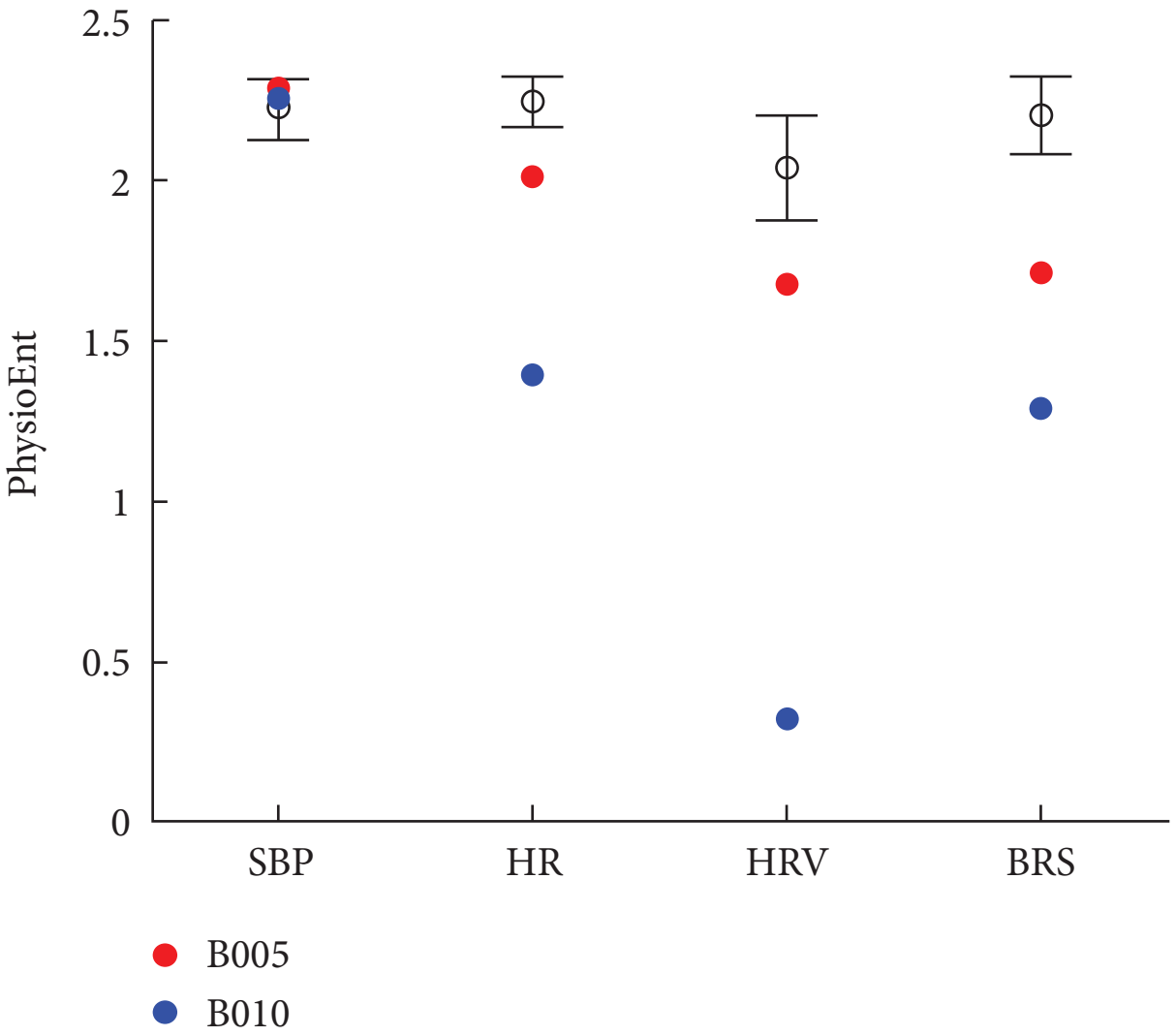

Figure 2: PhysioEnts for Eurobavar subjects. PhysioEnts of nonimpaired subjects are shown as means and error bars (means ± standard deviations). PhysioEnts of baroreflex-impaired subjects are remarked, respectively.

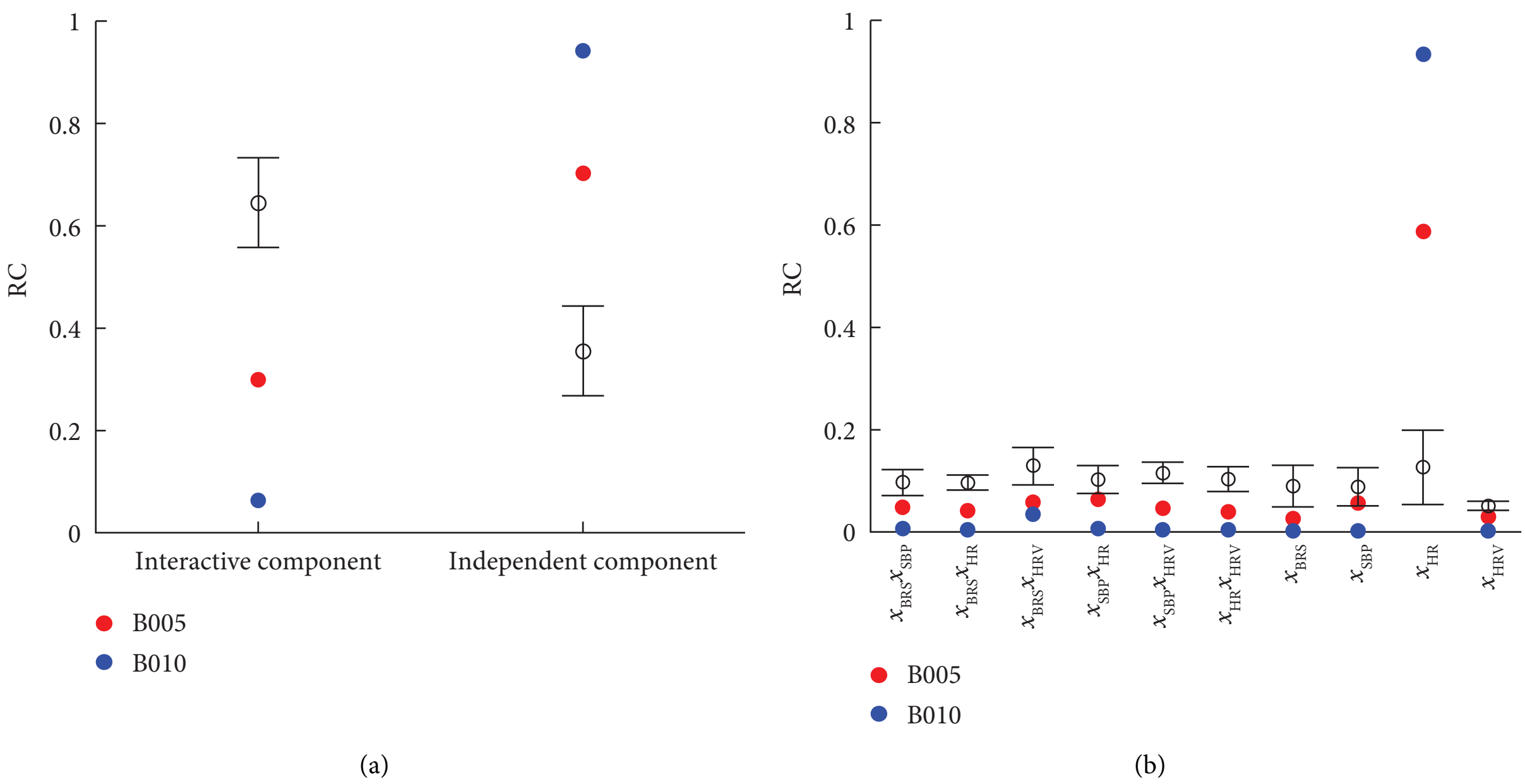

Figure 3: (a) The sum of interactive components' RCs and the sum of independent ones. (b) The RC of each model component.

Further, regression results in Table 4 show some significant correlations between demographic features and some model components' *RCs* (also, some of $p$ values are not great enough according to BH procedure but still may indicate some relatively significant correlations). Specifically, the *RC* of $x_{SBP}x_{HR}$ correlates positively with age, as depicted in Figure 11; the *RCs* of $x_{BRS}x_{HR}$, $x_{HR}x_{HRV}$, $x_{SBP}x_{HRV}$, and $x_{SBP}$ have significant associations with sex, as depicted in Figure 12; and the *RC* of $x_{BRS}$ correlates negatively with BMI, as provided in Figure 13. For the results in this section, further discussions are provided in section 6.2.2.

## 6. Discussion

*6.1. Resting-Condition BRF of Baroreflex-Impaired Subjects.* According to section 4, from Eurobavar dataset, the PhysioEnts and model components' *RCs* of baroreflex-impaired subjects are significantly differentiated with those of nonimpaired subjects. These may imply that:

Table 2: Multivariable linear regression results between PhysioEnts and demographic labels.

| Model components | PhysioEnt of SBP | PhysioEnt of HR | PhysioEnt of HRV | PhysioEnt of BRS |
|---|---|---|---|---|
| *Statistical results of F-test* | | | | |
| $F$ | 2.675 | 7.902 | 17.429 | 4.320 |
| $p^{F\text{-test}}$ | 0.0460[$] | 0.0000320* | $4.699E-11$* | 0.00488* |
| *Statistical results of t-test* | | | | |
| $\beta^*_{age}$ | 0.0766 | −0.0183 | −0.191 | −0.0936 |
| $\beta^*_{sex}$ | −0.0396 | 0.144 | 0.0560 | −0.0206 |
| $\beta^*_{BMI}$ | −0.0585 | −0.00133 | −0.0284 | 0.0934 |
| $t_{age}$ | 2.352 | −0.566 | −5.983 | −2.883 |
| $t_{sex}$ | −1.290 | 4.730 | 1.864 | −0.674 |
| $t_{BMI}$ | −1.785 | −0.0410 | −0.883 | 2.856 |
| $p^{t\text{-test}}_{age}$ | 0.0188* | 0.572 | $2.972E-9$* | 0.00401* |
| $p^{t\text{-test}}_{sex}$ | 0.197 | 0.000003* | 0.0626 | 0.500 |
| $p^{t\text{-test}}_{BMI}$ | 0.0745 | 0.967 | 0.377 | 0.00438* |

$F$ is the $F$ value of $F$-test for regressions. $p^{F\text{-test}}$ is the $p$ value of $F$-test. $\beta^*_{age}$, $\beta^*_{sex}$, and $\beta^*_{BMI}$ are standardized regression coefficients of age, sex, and BMI, respectively; $t_{age}$, $t_{sex}$, and $t_{BMI}$ are $t$ values of $t$-test for age, sex, and BMI; $p^{t\text{-test}}_{age}$, $p^{t\text{-test}}_{sex}$, and $p^{t\text{-test}}_{BMI}$ are $p$ values of $t$-test for age, sex, and BMI. *means that the $p$ value represents a significant result according to BH procedure. [$]means that the $p$ value is less than 0.05 but does not show significance according to BH procedure.

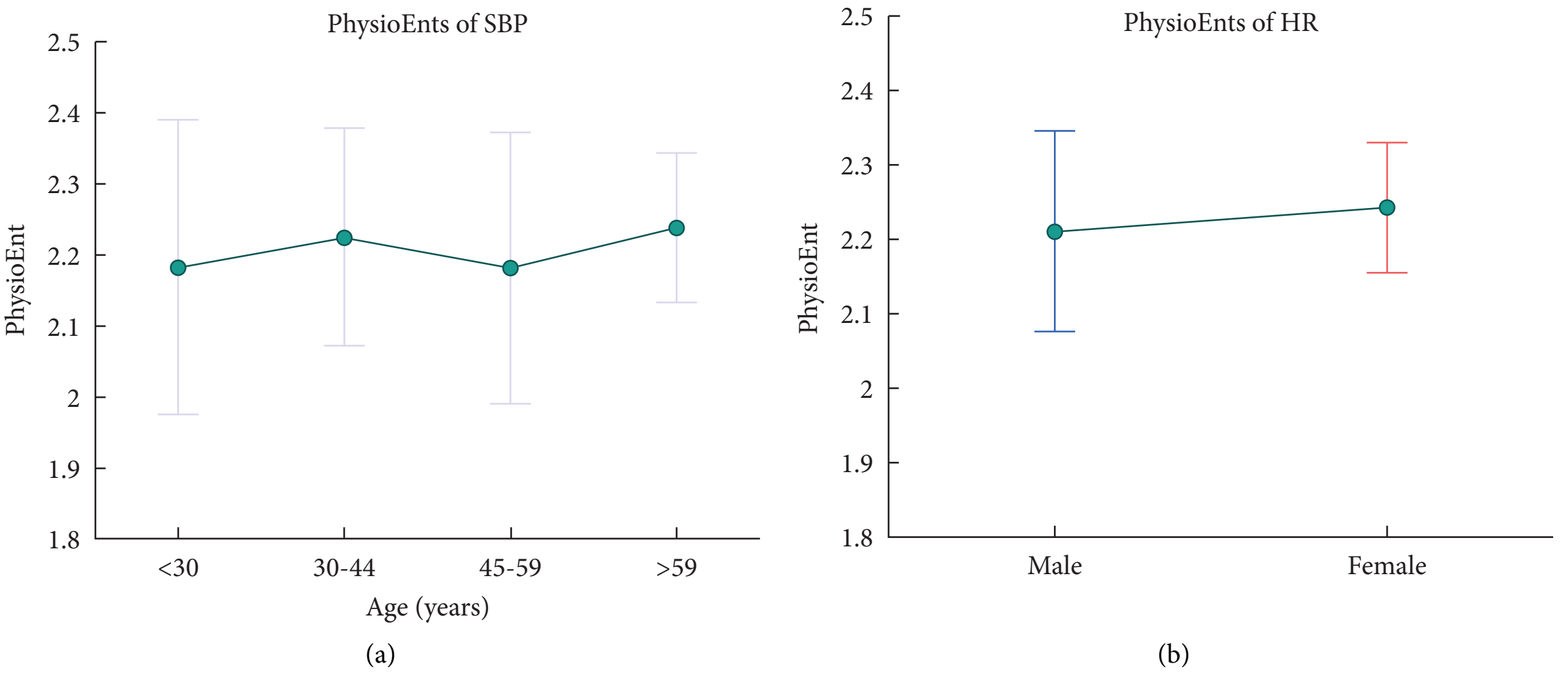

Figure 4: (a) Trend for the PhysioEnt of SBP with age. Subjects are further divided into four groups: young (<30 years), early middle age (30–44 years), middle age (45–59 years), and old (>59 years). (b) PhysioEnts of HR in males and females.

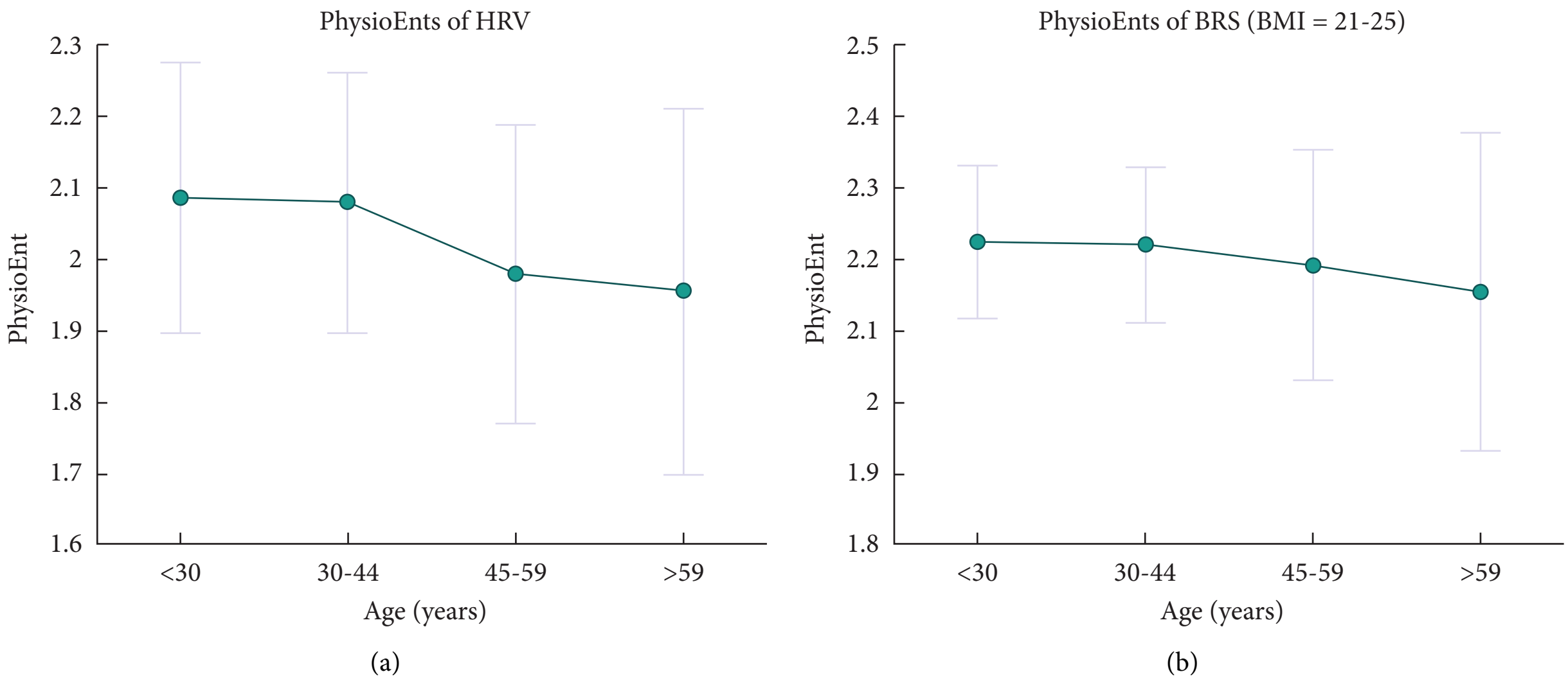

Figure 5: (a) Trend for the PhysioEnt of HRV with age, and subjects are further divided into four groups. (b) Trend for the PhysioEnt of BRS with BMI, in which the subjects with BMI 21–25 are depicted and divided into four age groups.

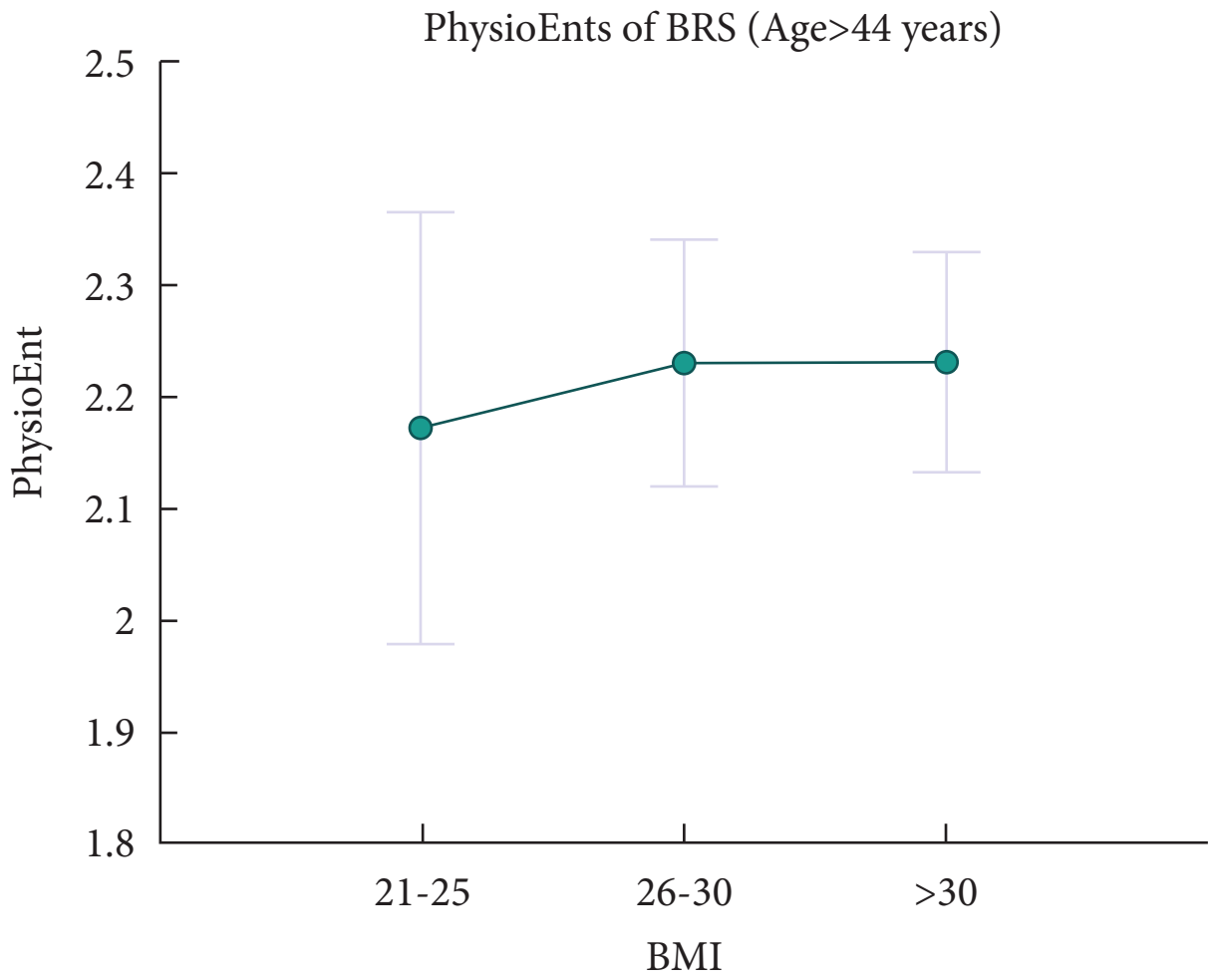

FIGURE 6: Trend for the PhysioEnt of BRS with BMI, in which the subjects with age >44 years are focused and divided into three BMI groups.

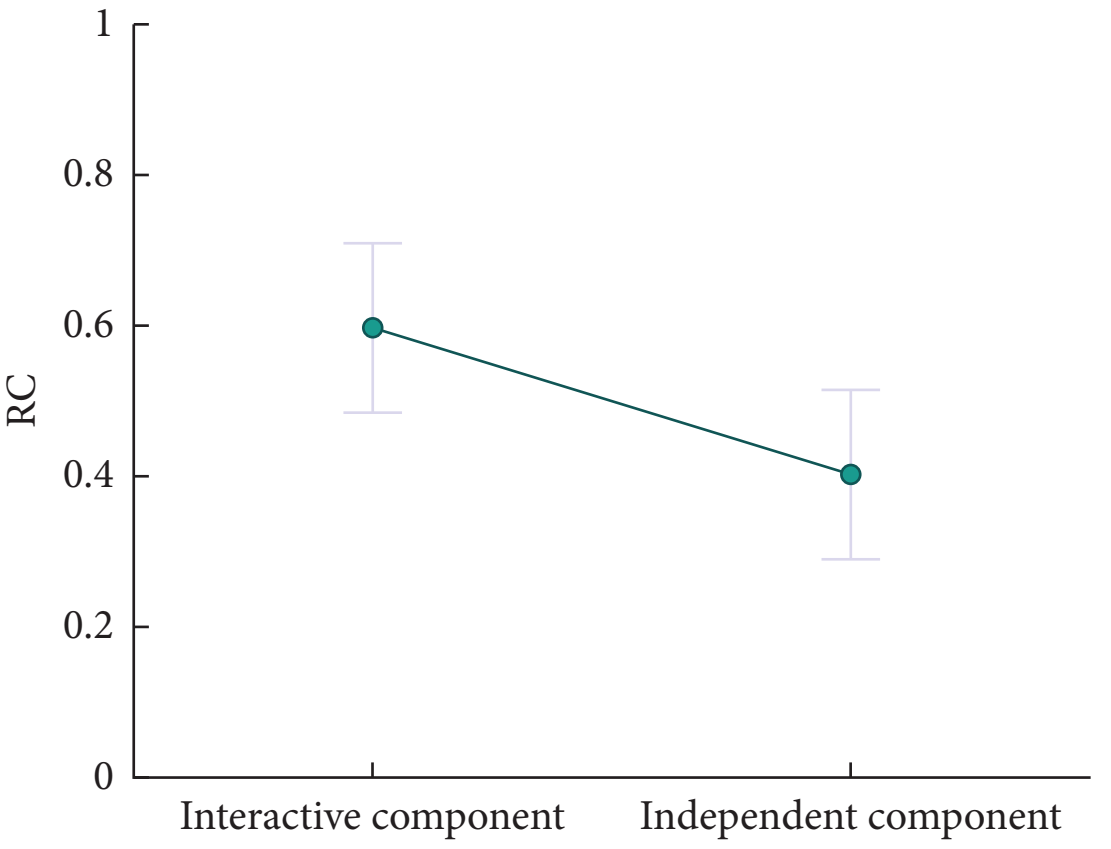

FIGURE 7: Comparison between the sum of interactive components' RCs and the sum of independent ones.

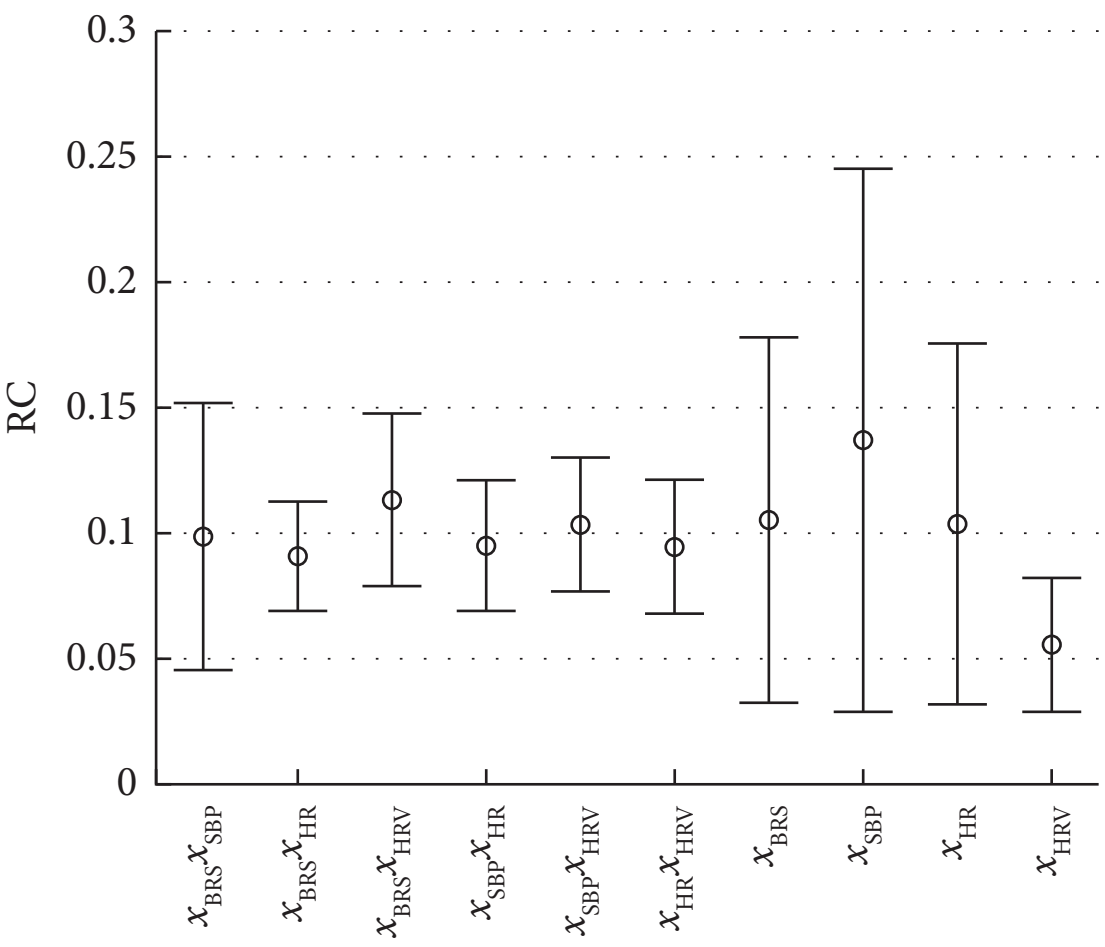

FIGURE 8: Comparison among model components' RCs.

(1) The PhysioEnts of HR, HRV, and BRS of impaired subjects are significantly low, which indicates the weaker responsiveness of effectors and neural activity according to section 3.1. Such a result corresponds to the worse health conditions of impaired subjects, and can support the rationality of the proposed method.

(2) The PhysioEnt of SBP of impaired subjects does not show a significant difference. Since subjects were tested in rest, there may not be stimuli great enough to disturb baroreflex, so that the fluctuations of SBP of different subjects cannot effectively reflect the difference in homeostasis health conditions.

(3) The *RC* of $x_{HR}$ is extremely greater for baroreflex-impaired subjects compared with the *RC*s of other model components, which indicates the significant independent contribution of HR to BRF. Such a result suggests that the impaired subjects' BRF is much sensitive to the function of effectors, i.e., sinus nodes. Further, their sinus nodes' function may be more independent and less correlated with other organs/tissues, which suggests that sinus nodes are more difficult to be driven and action. It has been found that baroreflex dysfunction can be due to the altered responsiveness of sinus node pacemaker cells to autonomic influences after heart diseases [69, 70], which may explain the weakened BRF for two subjects, especially for B010 undergoing a heart transplant procedure.

### *6.2. Demographic Features of the Resting-Condition BRF in Healthy Subjects*

*6.2.1. Demographic Features of PhysioEnts.* According to section 5.1, the demographic features of PhysioEnts may suggest that:

(1) Females own a greater PhysioEnt of HR than males, which indicates that females have a greater responsiveness of sinus nodes than males. For this point, current studies have found that females exhibited higher parasympathetic activity during rest than males [71, 72], which may explain such a sex difference in the responsiveness of sinus nodes.

(2) There are significant negative correlations between age and the PhysioEnts of HRV and BRS. Since the PhysioEnts of HRV and BRS represent the responsiveness of neural activity, such results suggest the weakened activity of parasympathetic nerves with aging. The degradation of parasympathetic nerves with aging has been found by existing studies [73, 74], which can support the above results concluded by the proposed method. Further, such parasympathetic neural deficits can contribute to age-associated reductions in BRF [75].

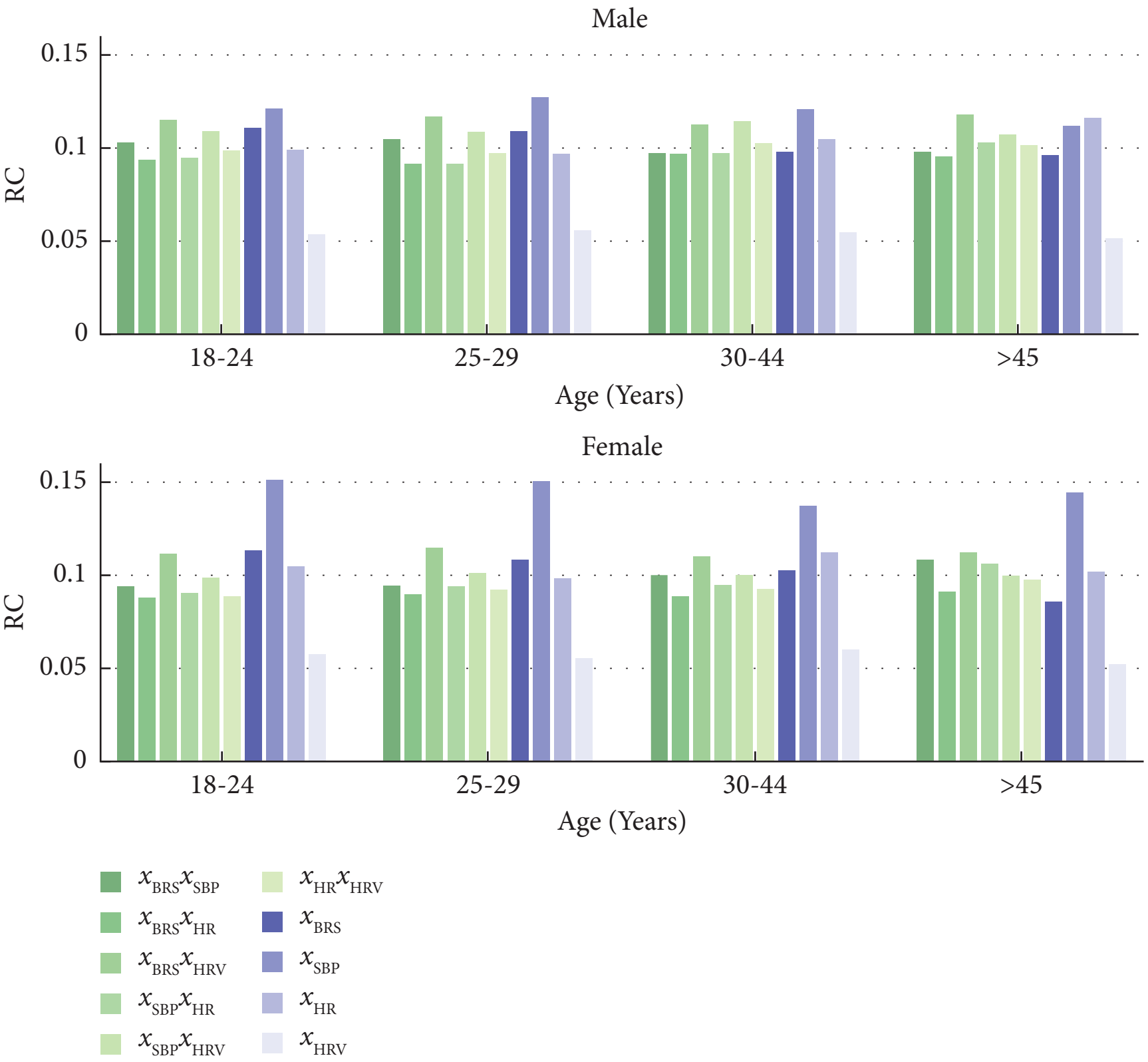

FIGURE 9: Comparisons among different model components' RCs in various groups. In each age and sex group, there are 10 boxes corresponding to the means of 10 model components' RCs.

TABLE 3: Multivariable linear regression results between the RCs' sums and demographic labels.

| Model components | The sum of interactive components | The sum of independent components |
|---|---|---|
| *Statistical results of F-test* | | |
| $F$ | 6.084 | 6.084 |
| $p^{F\text{-test}}$ | 0.000418* | 0.000418* |
| Statistical results of $t$-test | | |
| $\beta^*_{\text{age}}$ | 0.0572 | −0.0572 |
| $\beta^*_{\text{sex}}$ | −0.118 | 0.118 |
| $\beta^*_{\text{BMI}}$ | −0.0236 | 0.0236 |
| $t_{\text{age}}$ | 1.741 | −1.741 |
| $t_{\text{sex}}$ | −3.835 | 3.835 |
| $t_{\text{BMI}}$ | −0.713 | 0.713 |
| $p^{t\text{-test}}_{\text{age}}$ | 0.082 | 0.082 |
| $p^{t\text{-test}}_{\text{sex}}$ | 0.000133* | 0.000133* |
| $p^{t\text{-test}}_{\text{BMI}}$ | 0.476 | 0.476 |

*means that the $p$ value represents a significant result according to BH procedure.

(3) The PhysioEnt of BRS is positively correlated with BMI. It is considered that obesity has an adverse effect on BRF, which may be due to autonomic nervous system dysfunction [76]. Therefore, the positive correlation between the PhysioEnt of BRS and BMI still cannot be well explained, and may reflect some unclear physiological mechanisms.

(4) With a slight significance, the PhysioEnt of SBP correlates positively with age. According to section 3.1, a greater PhysioEnt of SBP may indicate a compromised internal homeostasis, which also suggests the worse resting-condition BRF of old people. In addition, similar to the analysis in section 6.1, the test condition in rest may influence how the fluctuation of SBP reflects the health condition of homeostasis, as well as the significance between the PhysioEnt of SBP and age.

*6.2.2. Analysis Related to Physiological Processes and Organs/Tissues.* According to section 5.2, there are some statistical significance on model components' *RC*s, which may reflect

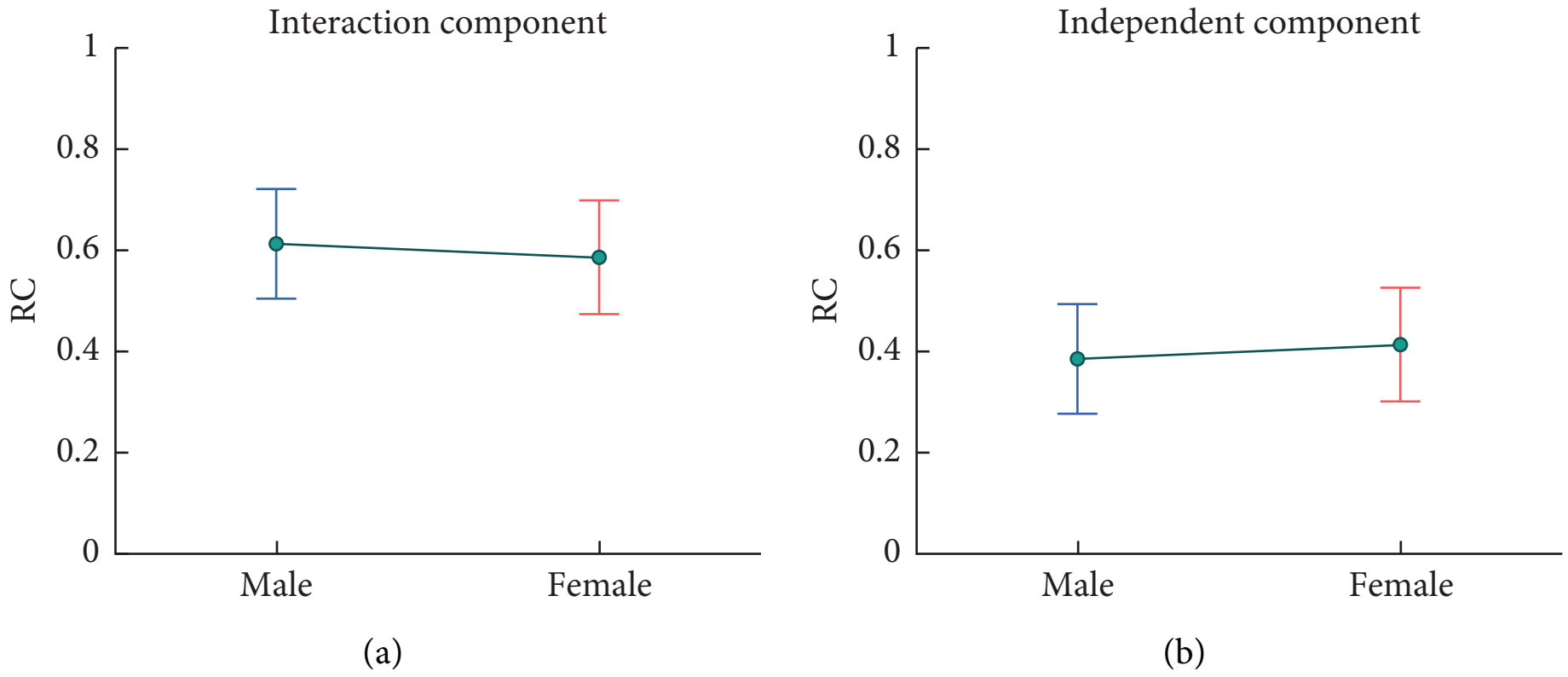

FIGURE 10: (a) The sum of interactive components' RCs in males and females. (b) The sum of independent components' RCs in males and females.

TABLE 4: Multivariable linear regression results between model components' RCs and demographic labels.

| Model component | $F$ | $p^{F\text{-test}}$ |
|---|---|---|
| *Statistical results of F-test* | | |
| $x_{\mathrm{BRS}}x_{\mathrm{SBP}}$ | 0.0603 | 0.981 |
| $x_{\mathrm{BRS}}x_{\mathrm{HR}}$ | 4.309 | 0.00496* |
| $x_{\mathrm{BRS}}x_{\mathrm{HRV}}$ | 1.155 | 0.326 |
| $x_{\mathrm{SBP}}x_{\mathrm{HR}}$ | 4.412 | 0.00430* |
| $x_{\mathrm{SBP}}x_{\mathrm{HRV}}$ | 8.396 | 0.0000160* |
| $x_{\mathrm{HR}}x_{\mathrm{HRV}}$ | 6.847 | 0.000143* |
| $x_{\mathrm{BRS}}$ | 3.126 | 0.0251$ |
| $x_{\mathrm{SBP}}$ | 2.662 | 0.0469$ |
| $x_{\mathrm{HR}}$ | 1.364 | 0.252 |
| $x_{\mathrm{HRV}}$ | 0.389 | 0.761 |

| Model component | $\beta^*_{\mathrm{age}}$ | $\beta^*_{\mathrm{sex}}$ | $\beta^*_{\mathrm{BMI}}$ |
|---|---|---|---|
| *Statistical results of t-test* | | | |
| $x_{\mathrm{BRS}}x_{\mathrm{SBP}}$ | 0.000859 | −0.0132 | −0.00184 |
| $x_{\mathrm{BRS}}x_{\mathrm{HR}}$ | 0.0240 | −0.107 | 0.00723 |
| $x_{\mathrm{BRS}}x_{\mathrm{HRV}}$ | 0.0265 | −0.0510 | −0.00826 |
| $x_{\mathrm{SBP}}x_{\mathrm{HR}}$ | 0.100 | −0.0396 | 0.00259 |
| $x_{\mathrm{SBP}}x_{\mathrm{HRV}}$ | 0.0283 | −0.151 | −0.0377 |
| $x_{\mathrm{HR}}x_{\mathrm{HRV}}$ | 0.0586 | −0.127 | −0.0433 |
| $x_{\mathrm{BRS}}$ | −0.0343 | 0.0249 | −0.0685 |
| $x_{\mathrm{SBP}}$ | −0.0226 | 0.0834 | 0.0331 |
| $x_{\mathrm{HR}}$ | −0.00979 | 0.0335 | 0.0600 |
| $x_{\mathrm{HRV}}$ | −0.0282 | −0.000110 | −0.0102 |

| Model component | $t_{\mathrm{age}}$ | $t_{\mathrm{sex}}$ | $t_{\mathrm{BMI}}$ | $p^{t\text{-test}}_{\mathrm{age}}$ | $p^{t\text{-test}}_{\mathrm{sex}}$ | $p^{t\text{-test}}_{\mathrm{BMI}}$ |
|---|---|---|---|---|---|---|
| $x_{\mathrm{BRS}}x_{\mathrm{SBP}}$ | 0.026 | −0.424 | −0.055 | 0.979 | 0.672 | 0.956 |
| $x_{\mathrm{BRS}}x_{\mathrm{HR}}$ | 0.729 | −3.462 | −0.218 | 0.466 | 0.000557* | 0.827 |
| $x_{\mathrm{BRS}}x_{\mathrm{HRV}}$ | 0.801 | −1.640 | −0.254 | 0.423 | 0.101 | 0.800 |
| $x_{\mathrm{SBP}}x_{\mathrm{HR}}$ | 3.045 | −1.280 | 0.0780 | 0.00239* | 0.201 | 0.938 |
| $x_{\mathrm{SBP}}x_{\mathrm{HRV}}$ | 0.863 | −4.919 | −1.144 | 0.388 | 0.000001* | 0.253 |
| $x_{\mathrm{HR}}x_{\mathrm{HRV}}$ | 1.785 | −4.123 | −1.309 | 0.0740 | 0.000040* | 0.191 |
| $x_{\mathrm{BRS}}$ | −1.041 | 0.803 | −2.064 | 0.298 | 0.422 | 0.0393$ |
| $x_{\mathrm{SBP}}$ | −0.684 | 2.692 | 0.997 | 0.494 | 0.00722* | 0.319 |
| $x_{\mathrm{HR}}$ | −0.296 | 1.077 | 1.801 | 0.767 | 0.282 | 0.0720 |
| $x_{\mathrm{HRV}}$ | −0.851 | −0.00355 | −0.305 | 0.395 | 0.997 | 0.761 |

*means that the $p$ value represents a significant result according to BH procedure. $means that the $p$ value is less than 0.05 but does not show significance according to BH procedure.

the features about physiological process and organs/tissues. Firstly, the overview for the comparison among components' *RC*s may imply some findings:

(1) The *RC* of $x_{\mathrm{BRS}}x_{\mathrm{HRV}}$ is the largest compared with other interactive components, which indicates that the interaction between BRS and HRV contributes most to the resting-condition BRF. Further, this may suggest that parasympathetic regulation affects the resting-condition BRF most significantly in all physiological processes. This result is consistent with the finding that some diseases related to parasympathetic nerves usually result in a decline of BRF [77–79].

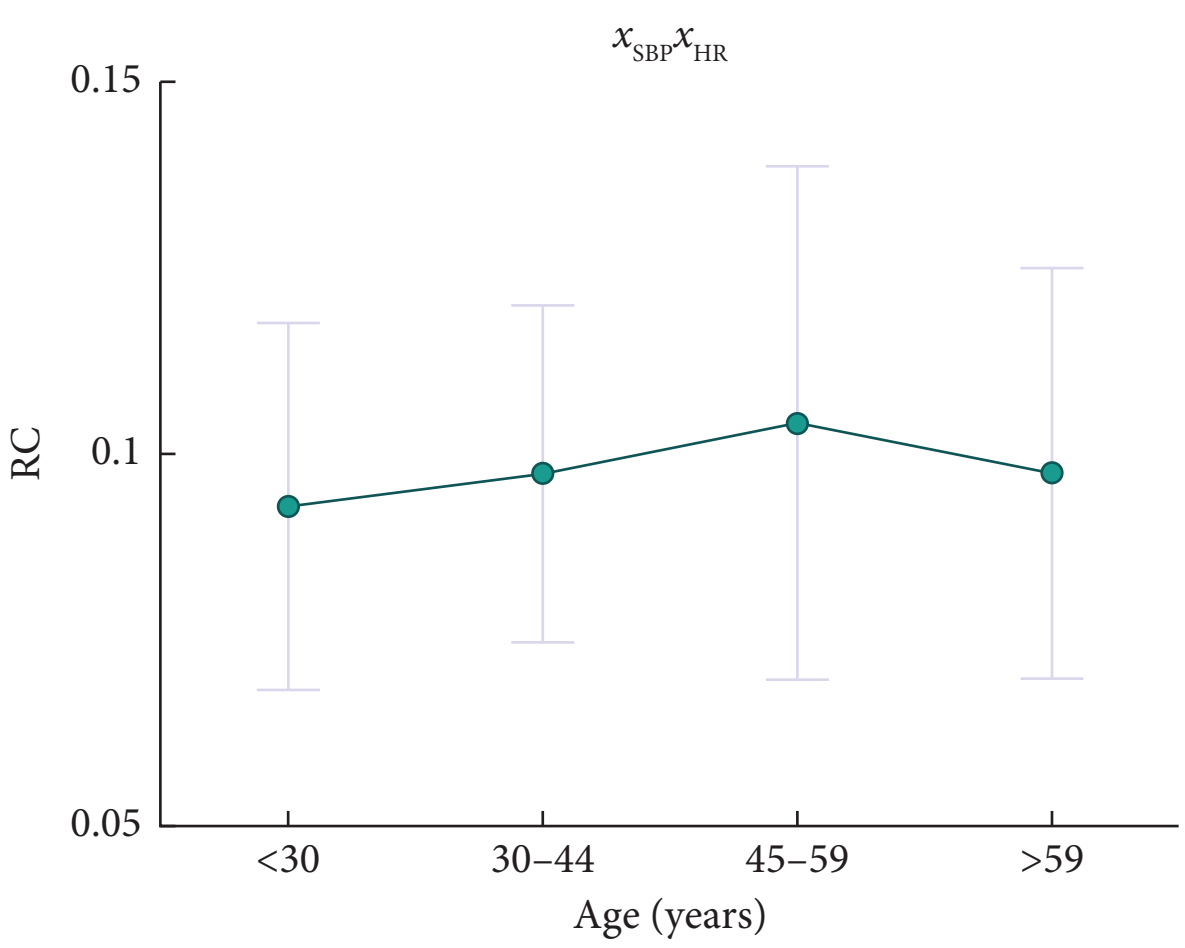

FIGURE 11: Trend for the RC of $x_{SBP}x_{HR}$ with age, in which the subjects are divided into four age groups.

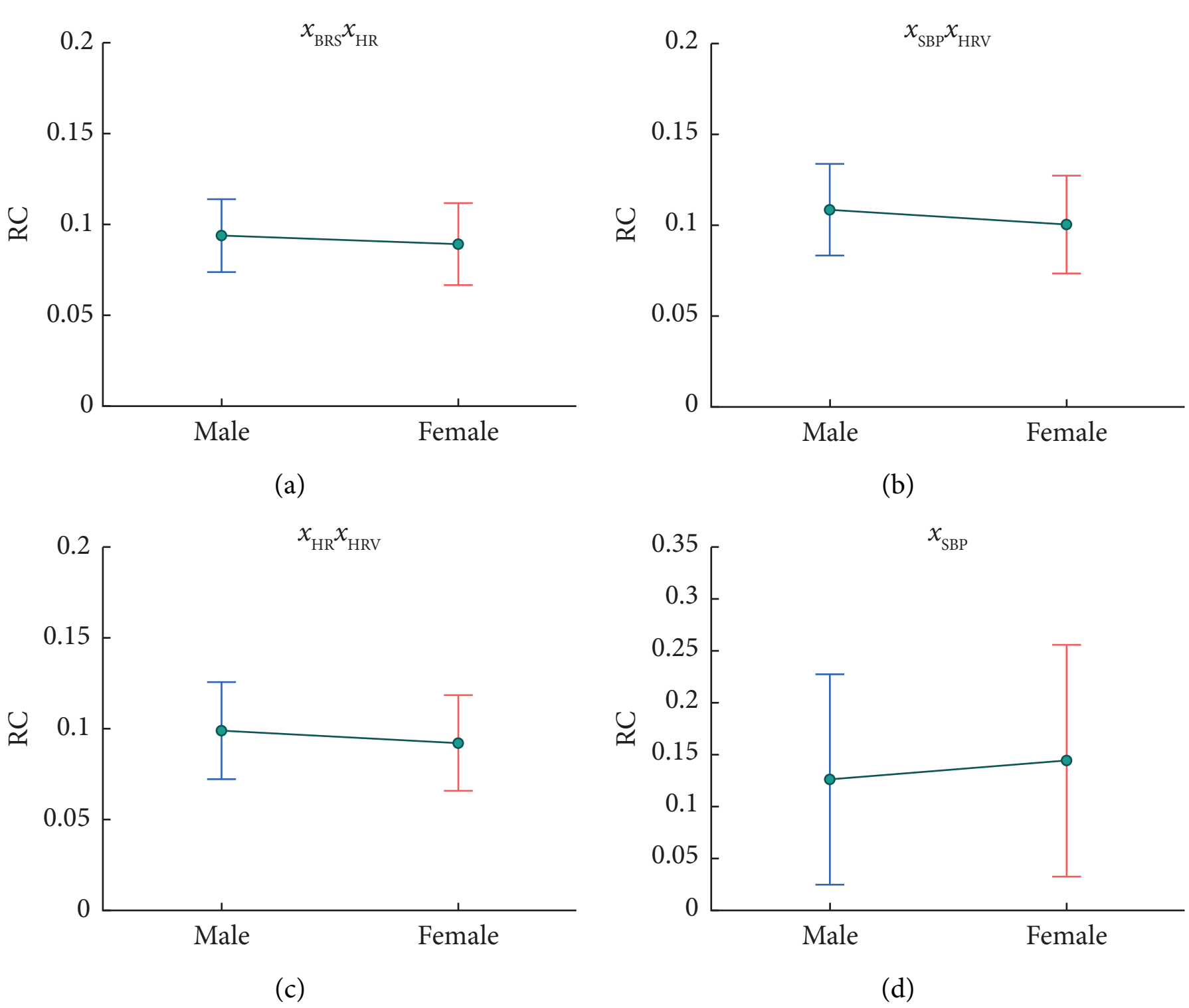

FIGURE 12: (a) The RC of $x_{BRS}x_{HR}$ in males and females. (b) The RC of $x_{HR}x_{HRV}$ in males and females. (c) The *RC* of $x_{SBP}x_{HRV}$ in males and females. (d) The *RC* of $x_{SBP}$ in males and females.

(2) The *RC* of $x_{SBP}$ is significantly large, which suggests that for healthy subjects, their resting-condition BRF is relatively sensitive to vascular characteristics. For this point, current studies have proved that the vascular mechanics of carotid sinuses and aortic arch, including vascular elasticity and resistance, are important contributors to BRF [80, 81], which can also support the critical contribution of blood vessels reflected by a great *RC* of $x_{SBP}$.

(3) The *RC* of $x_{HRV}$ is the smallest one. This indicates a relatively low independent contribution of HRV, which may reflect that parasympathetic nerves interact significantly with other organs/tissues. Further

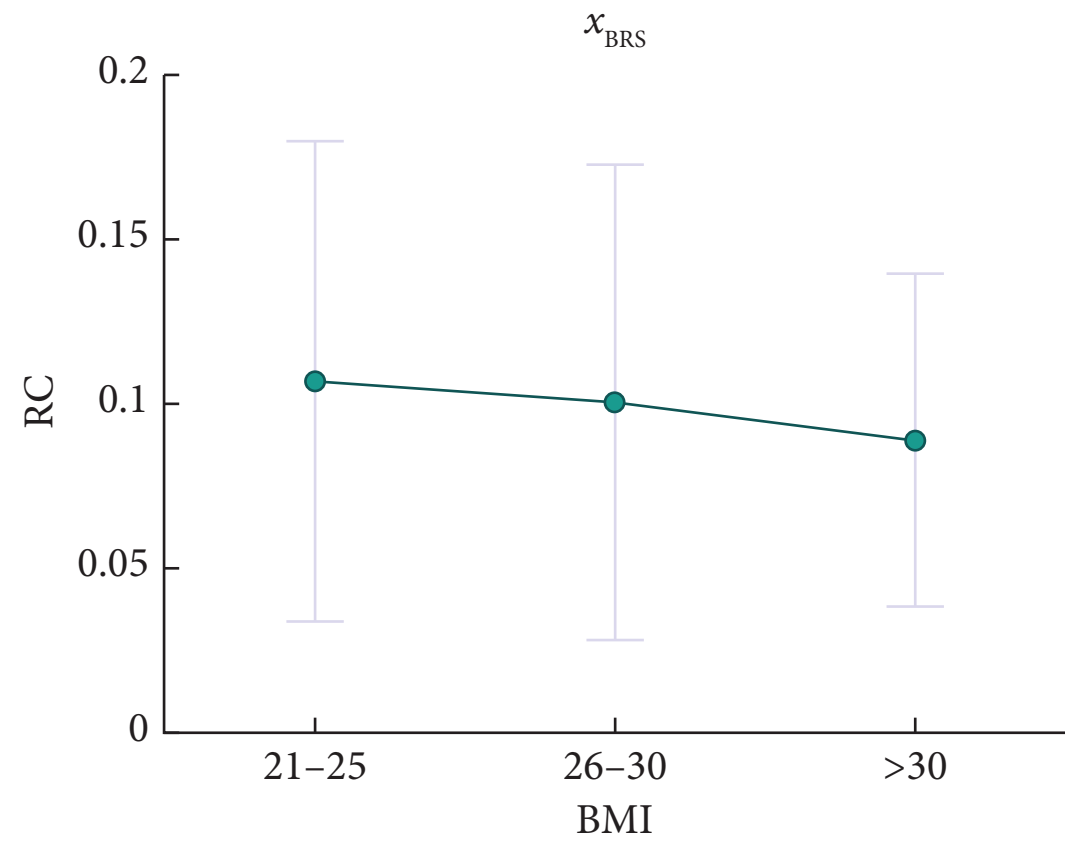

Figure 13: Trend for the RC of $x_{BRS}$ with BMI, in which the subjects are divided into three BMI groups.

studies on physiological mechanisms may be needed to explain this result.

In addition, there are correlations for some *RC*s with demographic features. For the correlation with age, the *RC* of $x_{SBP}x_{HR}$ increases with aging. This may suggest that older people's BRF is more sensitive to the process that HR regulates BP. Current studies have identified the degeneration of this process in old people [82, 83]. Therefore, it is more important for old people to maintain this process stable. Further, this can provide guidance for treatments to related diseases, and can also support some interventions for old patients aiming to enhance the regulation such as stimulating carotid sinuses [84] and parasympathetic nerves [85].

For sex differences, according to Table 3, the differences in interactive components and independent ones suggest that males depend more on physiological processes than females, while females depend more on the functions of organs/tissues than males. These results may explain sex differences in some baroreflex-related diseases. For example, it has been noted that males are more likely to die after a spontaneous intracerebral hemorrhage than females [86]. In addition, studies show that the weakening of BRF increases the risk of acute intracerebral hemorrhage and secondary brain injury [87]. According to this section, to keep BRF, for males, maintaining the functions of organs/tissues may not be as effective as for females. Therefore, when the failure of BRF occurs and leads to intracerebral hemorrhage, it may be more difficult for males to restore the homeostasis, which may be the explanation for a greater odd of dying in males.

Further, the *RC*s of $x_{BRS}x_{HR}$, $x_{HR}x_{HRV}$, $x_{SBP}x_{HRV}$, and $x_{SBP}$ have significant correlations with sex, which may suggest that:

(1) The larger *RC* of $x_{BRS}x_{HR}$ for males indicates that the interaction between BRS and HR contributes more to the resting-condition BRF. Since the relationship between short-term HR and BRS is not clear, the result suggests that males may depend more on related unclear physiological mechanisms.

(2) Males' *RC* of $x_{HR}x_{HRV}$ is larger, which implies that the interaction between HR and HRV contributes more to the resting-condition BRF. This may suggest males depend more on the process that parasympathetic nerves innervate HR. Some studies also reported that this process of males is less robust compared with the one of females [72, 88]. Therefore, it is more important for males to maintain this process stable.

(3) Males' *RC* of $x_{SBP}x_{HRV}$ is larger and indicates that the interaction between SBP and HRV contributes more for males. This may be an indirect reflection of the process related to parasympathetic nerves mentioned above.

(4) Females' *RC* of $x_{SBP}$ is larger, which represents a greater independent effect of SBP on the resting-condition BRF. Current studies also figured a greater contribution for vascular characteristics like aortic distensibility on BRF in females [80, 89], which may be due to the protective effect of estrogens on blood vessels [90].

For the feature related to BMI, the *RC* of $x_{BRS}$ shows a decrease with BMI, although the difference is not significant enough according to BH procedure. Such a result implies a less independent effect of BRS for obese subjects. Since $x_{BRS}$ represents the function of efferent parasympathetic nerves, this may suggest that the parasympathetic nerves disturbances of obesity people can reduce the dominance of nerves on BRF. And this phenomenon also has been recorded by current studies [91, 92].

*6.3. Rationality and Novelty of PhysioEnt.* In this section, the rationality and novelty of the proposed method are further discussed. The rationality of the proposed method is, essentially, the fluctuations of SBP, HR, HRV, and BRS can represent the health condition of BRF, which has been studied and proved by current physiological studies as stated in section 1. In fact, the general adopted BRS in clinic is also calculated based on the fluctuations of BP and HR. However, BRS has been proved to be partial when measuring BRF as mentioned in section 2.

Therefore, a systematic quantification of the index fluctuations is expected. In addition, and more importantly, we hope that this quantification is related to specific physiological mechanisms. For these purposes, a MaxEnt model from statistical physics is adopted, and PhysioEnts of four indexes can be calculated. According to the explanation in section 3.1 and current studies reviewed in section 1, it is also reasonable to model baroreflex regulation based on the principle of MaxEnt. From another side, the results and knowledge inferred based on the proposed method are consistent with the conclusions of many existing studies, as mentioned in section 6.1 and 6.2, which also support the rationality of the proposed method.

Further, compared with the traditional method quantifying BRF, like BRS, there are two major advances for PhysioEnts.

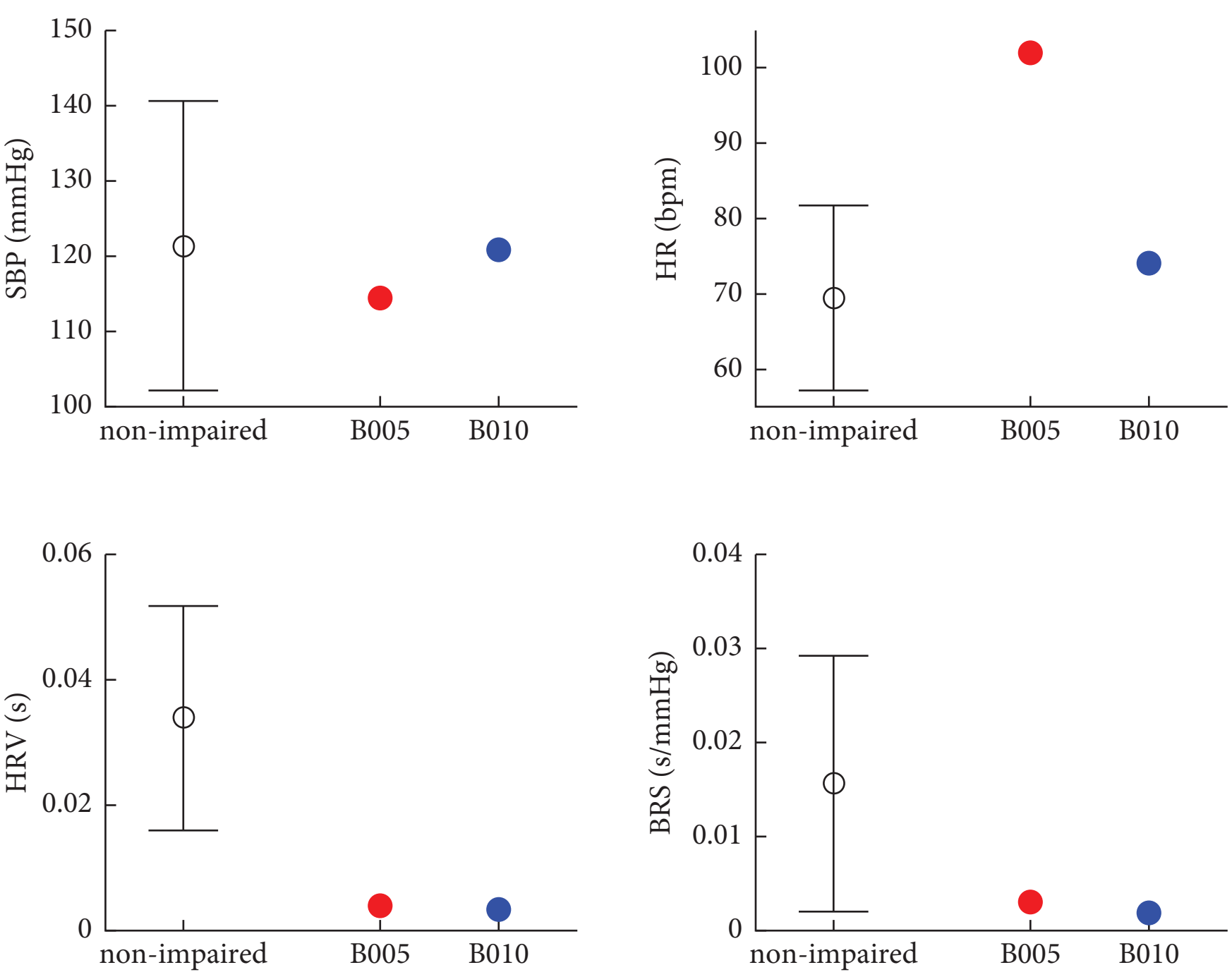

Figure 14: Means of SBP, HR, HRV, and HRV for Eurobavar subjects.

First, the PhysioEnts obtained by MaxEnt model incorporate more physiological indexes than just one BRS, which can capture more information about baroreflex. For this point, the features of single index are studied from Eurobavar dataset and Jena dataset to verify the necessity for MaxEnt model to combine indexes. Result from Eurobavar subjects is provided in Figure 14, which shows means of physiological indexes.

From Figure 13, it indicates that the HRV and BRS of impaired subjects are significantly lower than those of nonimpaired subjects, which suggests the compromised parasympathetic nerves. However, such levels of indexes cannot further discuss the responsiveness of effectors and neural activity. Also, it cannot support to infer the contribution and status of patients' effectors as mentioned in section 6.1.

For the healthy subjects of Jena dataset, the correlation for each index with age, sex, and BMI is also studied. Statistical results are depicted in Table 5.

Results show that there are only negative correlations for BRS and HRV with age, and significant sex differences for SBP and HR can be cognized. Both BRS and HRV represent the function of autonomic nervous system. Especially for BRS, it has been generally adopted to measure BRF in clinic. However, it should be figured that the fluctuations but not the levels of SBP and HR, are the basis to measure BRF [67]. Therefore, the individual analysis on BRS and HRV can only find the influence of aging, but not the effect of obesity and sex. By comparison, according to section 6.2, the MaxEnt model and corresponding PhysioEnts can capture more effective information about the resting-condition BRF.

Besides the necessity to combine indexes, more importantly, as mentioned before, we expect to achieve a brief and effective systematic characterization of complex physiological mechanisms when quantifying the resting-condition BRF. Traditional quantifications of BRF, such as BRS, are generally the observation and calculation for changes of BP and HR. Such quantifications are essentially black-box methods, which lack the description of physiological mechanisms. Therefore, for the statistical analysis of indexes including BRS, it is difficult to infer the factors affecting BRF, not even to help for improving BRF. On the other hand, the detailed modeling and analysis on the mechanisms of baroreflex are also difficult, and a complete white-box model could be complicated and unrealistic in clinical application.

To solve these problems, as mentioned in section 3.1, the proposed method constructs a hypothetical MaxEnt model based on physiological mechanisms in advance, and complete the model based on observed data. When building the model, each model component corresponds to some specific part of baroreflex regulation in advance. Therefore, the MaxEnt model and calculated PhysioEnts can not only represent BRF, but also study the effects of specific organs/tissues and physiological processes on BRF based on the *RC* proposed in section 3.2. Conclusively, the MaxEnt model and PhysioEnts are expected to reflect the emergence from the bottom-layer physiological mechanisms to the middle-layer physiological indexes and the top-layer BRF, as stated in section 2.

According to the above advances, the proposed method can capture some novel conclusions compared with traditional indexes like BRS. A typical instance is the

Table 5: Multivariable linear regression results between each index and demographic labels.

| Physiological index | $F$ | $p^{F\text{-test}}$ |
|---|---|---|
| *Statistical results of F-test* | | |
| BRS | 10.264 | <1$E$ − 10* |
| HRV | 14.075 | <1$E$ − 10* |
| SBP | 4.080 | 0.007* |
| HR | 7.546 | 0.000054* |

| Model component | $\beta^*_{\text{age}}$ | $\beta^*_{\text{sex}}$ | $\beta^*_{\text{BMI}}$ |
|---|---|---|---|
| *Statistical results of t-test* | | | |
| BRS | −0.171 | 0.009 | 0.010 |
| HRV | −0.168 | 0.056 | −0.034 |
| SBP | 0.044 | 0.099 | −0.023 |
| HR | 0.055 | 0.134 | 0.030 |

| Model component | $t_{\text{age}}$ | $p^{t\text{-test}}_{\text{age}}$ | $t_{\text{sex}}$ | $p^{t\text{-test}}_{\text{sex}}$ | $t_{\text{BMI}}$ | $p^{t\text{-test}}_{\text{BMI}}$ |
|---|---|---|---|---|---|---|
| BRS | −5.245 | 1.88$E$ − 7* | 0.288 | 0.773 | 0.316 | 0.752 |
| HRV | −5.173 | 2.76$E$ − 7* | 1.833 | 0.067 | −1.037 | 0.300 |
| SBP | 1.340 | 0.181 | 3.189 | 0.001* | −0.699 | 0.485 |
| HR | 1.674 | 0.094 | 4.364 | 0.000014* | 0.919 | 0.358 |

*means that the $p$ value represents a significant result according to BH procedure.

contributions of organs/tissues and physiological processes on BRF. According to section 6.2.2, the physiological processes among organs/tissues are more critical for the resting-condition BRF, especially for males. On the one hand, this may be corroborated by some clinical findings, such as the unclear high mortality for males after a spontaneous intracerebral hemorrhage, which may result from the fact that males are more difficult to recover from regulation failure due to the stronger dependencies among organs/tissues. On the other hand, maintaining physiological processes is more meaningful to recover and strengthen BRF. It should be noted that such analyses that how specific physiological mechanisms affect BRF cannot be obtained from single BRS.

*6.4. Limitations.* In this work, a MaxEnt model with physiological indexes is established, and the PhysioEnt is proposed to quantify the resting-condition BRF. Some new medical findings are also obtained only based on limited data. Consequently, there are some limitations existing in the work, mainly including the integrity of modeling and data sufficiency.

For MaxEnt modeling, four physiological indexes are selected based on current medical knowledge and testability. On the one hand, the assumption for stable sympathetic neural activity and the usage of four indexes to quantify the resting-condition BRF may not perfectly fit the real baroreflex; on the other hand, the HRV is represented indirectly by standard deviation of RRI or IBI, and the BRS quantified in frequency domain also may not be precise indexes. For this point, complementary methods to identify physiological mechanisms and evaluate indexes should be further studied and incorporated.

For data sufficiency, the effects of more physiological systems such as the renin-angiotensin-aldosterone system (RAAS) cannot be considered due to the lack of data. In addition, the sample sizes of adopted datasets are relatively limited, which may also affect the accuracy of conclusions.

## 7. Conclusion

To achieve a systematic understanding on the function of complex baroreflex system based on measurable indexes, in this work, the entropy is adopted to quantify the resting-condition baroreflex function (BRF), and the maximum entropy approach is applied to model physiological interactions. Two open-source datasets are analyzed, and the following conclusions can be drawn, which are consistent with current studies and support the rationality of the proposed method.

(1) Compared with nonimpaired subjects, the baroreflex-impaired patients possess significantly weaker responsiveness of sinus nodes and neural activity

(2) For healthy subjects, there is a degradation of parasympathetic nerves and a weakening of BRF with aging

(3) A sex difference exists in healthy subjects that females exhibit higher parasympathetic activity and better responsiveness of sinus nodes

Besides the above conclusions, some novel insights also can be provided, which can be further studied and discussed. For instance, there is a positive correlation between the fluctuation of BRS and BMI; additionally, compared with females, males may rely more on the reliable physiological interactions among organs/tissues.

For the future work, further studies and clinical trials can be performed according to the findings and conclusions in this work. In addition, based on the proposed paradigm, complementary physiological methods will be included to improve models, and more physiological systems that affect BRF will be concerned with proper datasets. Meanwhile, the proposed method can be generalized beyond resting condition using individualized dynamic monitoring data in daily life. Prospectively, this work can support the individualized real-time evaluation for healthy status of BRF and guide precision medicine.

## Data Availability

Eurobavar dataset is available via https://www.eurobavar.altervista.org/eurobavar.html Jena dataset is available via https://physionet.org/content/autonomic-aging-cardiovascular/1.0.0/.

## Disclosure

This work has been submitted as a pre-print in [93], https://arxiv.org/abs/2308.14584.

## Conflicts of Interest

The authors declare that they have no conflicts of interest.

## Acknowledgments

This work was supported by the National Natural Science Foundation of China (Grant no. 51775020), the National Natural Science Foundation of China (Grant no. 62073009), and the Beijing Natural Science Foundation of China (Grant no. 7222086).

## Supplementary Materials

Supplementary materials include: Figure S1. Monitoring data for one subject in Eurobavar dataset; Figure S2. Monitoring data for one subject in Jena dataset; Table S1. Expectation differences in the *RC*s of every two model components according to two-tailed Games–Howell tests; Table S2. $p$ values of two-tailed Games–Howell tests for every two model components; Supplementary Texts: (1) Analysis on the model components in $P_{\mathrm{II}}(x)$; (2) Analysis on the data discretization, which includes the comparisons between the statistical results with different discretized levels: Table S3. Multivariable linear regression results for the PhysioEnt of SBP with different discretized levels; Table S4. Multivariable linear regression results for the PhysioEnt of HR with different discretized levels; Table S5. Multivariable linear regression results for the PhysioEnt of HRV with different discretized levels; Table S6. Multivariable linear regression results for the PhysioEnt of BRS with different discretized levels. (*Supplementary Materials*)

## References

[1] F. Gu, E. B. Randall, S. Whitesall et al., "Potential role of intermittent functioning of baroreflexes in the etiology of hypertension in spontaneously hypertensive rats," *JCI insight*, vol. 5, no. 19, Article ID e139789, 2020.

[2] S. Tang, L. Xiong, Y. Fan, V. C. Mok, K. S. Wong, and T. W. Leung, "Stroke outcome prediction by blood pressure variability, heart rate variability, and baroreflex sensitivity," *Stroke*, vol. 51, no. 4, pp. 1317–1320, 2020.

[3] M. T. L. Rovere, J. T. Bigger, F. I. Marcus, A. Mortara, P. J. Schwartz, and A. Investigators, "Baroreflex sensitivity and heart-rate variability in prediction of total cardiac mortality after myocardial infarction," *The Lancet*, vol. 351, no. 9101, pp. 478–484, 1998.

[4] S. L. Bealer, "Peripheral hyperosmolality reduces cardiac baroreflex sensitivity," *Autonomic Neuroscience*, vol. 104, no. 1, pp. 25–31, 2003.

[5] S. Ogoh, R. M. Brothers, Q. Barnes et al., "Effects of changes in central blood volume on carotid-vasomotor baroreflex sensitivity at rest and during exercise," *Journal of Applied Physiology*, vol. 101, no. 1, pp. 68–75, 2006.

[6] B. Kent, J. Drane, B. Blumenstein, and J. Manning, "A mathematical model to assess changes in the baroreceptor reflex," *Cardiology*, vol. 57, no. 5, pp. 295–310, 1972.

[7] M. Zamir, M. B. Badrov, T. D. Olver, and J. K. Shoemaker, "Cardiac baroreflex variability and resetting during sustained mild effort," *Frontiers in Physiology*, vol. 8, p. 246, 2017.

[8] K. Kamada, K. Saku, T. Tohyama et al., "Diabetes mellitus attenuates the pressure response against hypotensive stress by impairing the sympathetic regulation of the baroreflex afferent arc," *American Journal of Physiology-Heart and Circulatory Physiology*, vol. 316, no. 1, pp. H35–H44, 2019.

[9] M. Moslehpour, T. Kawada, K. Sunagawa, M. Sugimachi, and R. Mukkamala, "Nonlinear identification of the total baroreflex arc: chronic hypertension model," *American Journal of Physiology - Regulatory, Integrative and Comparative Physiology*, vol. 310, no. 9, pp. R819–R827, 2016.

[10] C. A. Swenne, "Baroreflex sensitivity: mechanisms and measurement," *Netherlands Heart Journal*, vol. 21, no. 2, pp. 58–60, 2013.

[11] I. Biaggioni, C. A. Shibao, and J. Jordan, "Evaluation and diagnosis of afferent baroreflex failure," *Hypertension*, vol. 79, no. 1, pp. 57–59, 2022.

[12] P. Ataee, G. A. Dumont, H. A. Noubari, W. T. Boyce, and J. M. Ansermino, "A novel approach to the design of an artificial bionic baroreflex," in *Presented at 2013 35th Annual International Conference of the IEEE Engineering in Medicine and Biology Society (EMBC)*, pp. 3813–3816, IEEE, March 2013.

[13] K. D. Lau and C. A. Figueroa, "Simulation of short-term pressure regulation during the tilt test in a coupled 3d–0d closed-loop model of the circulation," *Biomechanics and Modeling in Mechanobiology*, vol. 14, no. 4, pp. 915–929, 2015.

[14] D. Canuto, K. Chong, C. Bowles, E. P. Dutson, J. D. Eldredge, and P. Benharash, "A regulated multiscale closed-loop cardiovascular model, with applications to hemorrhage and hypertension," *International Journal for Numerical Methods in Biomedical Engineering*, vol. 34, no. 6, Article ID e2975, 2018.

[15] T. Raphan, B. Cohen, Y. Xiang, and S. B. Yakushin, "A model of blood pressure, heart rate, and vaso-vagal responses produced by vestibulo-sympathetic activation," *Frontiers in Neuroscience*, vol. 10, p. 96, 2016.

[16] Y. M. Ishbulatov, A. S. Karavaev, A. R. Kiselev et al., "Mathematical modeling of the cardiovascular autonomic control in healthy subjects during a passive head-up tilt test," *Scientific Reports*, vol. 10, pp. 16525–16611, 2020.

[17] R. Bighamian, B. Parvinian, C. G. Scully, G. Kramer, and J.-O. Hahn, "Control-oriented physiological modeling of hemodynamic responses to blood volume perturbation," *Control Engineering Practice*, vol. 73, pp. 149–160, 2018.

[18] L. Fresiello, B. Meyns, A. Di Molfetta, and G. Ferrari, "A model of the cardiorespiratory response to aerobic exercise in healthy and heart failure conditions," *Frontiers in Physiology*, vol. 7, p. 189, 2016.

[19] L. Fresiello, A. W. Khir, A. Di Molfetta, M. Kozarski, and G. Ferrari, "Effects of intra-aortic balloon pump timing on baroreflex activities in a closed-loop cardiovascular hybrid model," *Artificial Organs*, vol. 37, no. 3, pp. 237–247, 2013.

[20] P. E. Hammer and J. P. Saul, "Resonance in a mathematical model of baroreflex control: arterial blood pressure waves accompanying postural stress," *American Journal of Physiology - Regulatory, Integrative and Comparative Physiology*, vol. 288, no. 6, pp. R1637–R1648, 2005, https://www.ncbi.nlm.nih.gov/pubmed/15718393.

[21] M. Di Rienzo, G. Mancia, and G. Parati, *Blood Pressure and Heart Rate Variability: Computer Analysis, Modelling and Clinical Applications*, IOs press, 1993.

[22] R. D. Berger, J. P. Saul, and R. J. Cohen, "Transfer function analysis of autonomic regulation. I. Canine atrial rate response," *American Journal of Physiology - Heart and Circulatory Physiology*, vol. 256, no. 1, pp. H142–H152, 1989.

[23] A. Porta, A. Marchi, V. Bari et al., "Assessing the strength of cardiac and sympathetic baroreflex controls via transfer entropy during orthostatic challenge," *Philosophical Transactions of the Royal Society A: Mathematical, Physical & Engineering Sciences*, vol. 375, no. 2096, Article ID 20160290, 2017.

[24] W.-M. Liu, H.-R. Liu, P.-W. Chen, H.-R. Chang, C.-M. Liao, and A.-B. Liu, "Novel application of multiscale cross-approximate entropy for assessing early changes in the complexity between systolic blood pressure and ecg rr intervals in diabetic rats," *Entropy*, vol. 24, no. 4, p. 473, 2022.

[25] C. Diao and N. Cai, "Temporal variation measure analysis: an improved second-order difference plot," *Complexity*, vol. 2022, no. 1, 2022.

[26] B. Wang, D. Liu, X. Gao, and Y. Luo, "Three-dimensional poincaré plot analysis for heart rate variability," *Complexity*, vol. 2022, no. 1, p. 2022, 2022.

[27] A. Barajas-Martínez, E. Ibarra-Coronado, M. P. Sierra-Vargas et al., "Physiological network from anthropometric and blood test biomarkers," *Frontiers in Physiology*, vol. 11, Article ID 612598, 2020.

[28] L. Cloarec-Blanchard, "Heart rate and blood pressure variability in cardiac diseases: pharmacological implications," *Fundamental & Clinical Pharmacology*, vol. 11, no. 1, pp. 19–28, 1997.

[29] F. Nikolaou, C. Orphanidou, P. Papakyriakou, K. Murphy, R. G. Wise, and G. D. Mitsis, "Spontaneous physiological variability modulates dynamic functional connectivity in resting-state functional magnetic resonance imaging," *Philosophical Transactions of the Royal Society A: Mathematical, Physical & Engineering Sciences*, vol. 374, no. 2067, Article ID 20150183, 2016.

[30] K. Hu, P. C. Ivanov, Z. Chen, M. F. Hilton, H. E. Stanley, and S. A. Shea, "Non-random fluctuations and multi-scale dynamics regulation of human activity," *Physica A: Statistical Mechanics and Its Applications*, vol. 337, no. 1-2, pp. 307–318, 2004.

[31] L. Nunes Amaral, P. C. Ivanov, N. Aoyagi et al., "Behavioral-independent features of complex heartbeat dynamics," *Physical Review Letters*, vol. 86, no. 26, pp. 6026–6029, 2001.

[32] R. Fossion, A. L. Rivera, and B. Estanol, "A physicist's view of homeostasis: how time series of continuous monitoring reflect the function of physiological variables in regulatory mechanisms," *Physiological Measurement*, vol. 39, no. 8, Article ID 084007, 2018.

[33] G. Parati, J. E. Ochoa, C. Lombardi, and G. Bilo, "Blood pressure variability: assessment, predictive value, and potential as a therapeutic target," *Current Hypertension Reports*, vol. 17, no. 4, pp. 23–18, 2015.

[34] M. Meyer and O. Stiedl, "Self-affine fractal variability of human heartbeat interval dynamics in health and disease," *European Journal of Applied Physiology*, vol. 90, no. 3-4, pp. 305–316, 2003.

[35] B. R. Pittman-Polletta, F. A. Scheer, M. P. Butler, S. A. Shea, and K. Hu, "The role of the circadian system in fractal neurophysiological control," *Biological Reviews*, vol. 88, no. 4, pp. 873–894, 2013.

[36] T. Balch, "Hierarchic social entropy: an information theoretic measure of robot group diversity," *Autonomous Robots*, vol. 8, no. 3, pp. 209–238, 2000.

[37] D. Zachary and S. Dobson, "Urban development and complexity: Shannon entropy as a measure of diversity," *Planning Practice and Research*, vol. 36, no. 2, pp. 157–173, 2021.

[38] T. Mora, A. M. Walczak, W. Bialek, and C. G. Callan, "Maximum entropy models for antibody diversity," *Proceedings of the National Academy of Sciences of the United States of America*, vol. 107, no. 12, pp. 5405–5410, 2010, https://www.ncbi.nlm.nih.gov/pubmed/20212159.

[39] E. T. Jaynes, "Information theory and statistical mechanics," *Physical Review*, vol. 106, no. 4, pp. 620–630, 1957.

[40] E. T. Jaynes, "Information theory and statistical mechanics. Ii," *Physical Review*, vol. 108, no. 2, pp. 171–190, 1957.

[41] W. Bialek, A. Cavagna, I. Giardina et al., "Statistical mechanics for natural flocks of birds," *Proceedings of the National Academy of Sciences*, vol. 109, no. 13, pp. 4786–4791, 2012.

[42] T. R. Lezon, J. R. Banavar, M. Cieplak, A. Maritan, and N. V. Fedoroff, "Using the principle of entropy maximization to infer genetic interaction networks from gene expression patterns," *Proceedings of the National Academy of Sciences*, vol. 103, no. 50, pp. 19033–19038, 2006.

[43] E. Agliari, F. Alemanno, A. Barra et al., "Analysis of temporal correlation in heart rate variability through maximum entropy principle in a minimal pairwise glassy model," *Scientific Reports*, vol. 10, no. 1, Article ID 15353, 2020.

[44] T. Ezaki, T. Watanabe, M. Ohzeki, and N. Masuda, "Energy landscape analysis of neuroimaging data," *Philosophical Transactions of the Royal Society A: Mathematical, Physical & Engineering Sciences*, vol. 375, no. 2096, Article ID 20160287, 2017.

[45] S. Gu, M. Cieslak, B. Baird et al., "The energy landscape of neurophysiological activity implicit in brain network structure," *Scientific Reports*, vol. 8, no. 1, p. 2507, 2018.

[46] R. Perini and A. Veicsteinas, "Heart rate variability and autonomic activity at rest and during exercise in various physiological conditions," *European Journal of Applied Physiology*, vol. 90, no. 3-4, pp. 317–325, 2003.

[47] J. F. Thayer and E. Sternberg, "Beyond heart rate variability: vagal regulation of allostatic systems," *Annals of the New York Academy of Sciences*, vol. 1088, no. 1, pp. 361–372, 2006.

[48] B. G. Wallin and N. Charkoudian, "Sympathetic neural control of integrated cardiovascular function: insights from measurement of human sympathetic nerve activity," *Muscle & Nerve*, vol. 36, no. 5, pp. 595–614, 2007.

[49] G. Sundlöf and B. Wallin, "Human muscle nerve sympathetic activity at rest. Relationship to blood pressure and age," *Journal of Physiology*, vol. 274, no. 1, pp. 621–637, 1978.

[50] J. Fagius and B. G. Wallin, "Long-term variability and reproducibility of resting human muscle nerve sympathetic activity at rest, as reassessed after a decade," *Clinical Autonomic Research*, vol. 3, pp. 201–205, 1993.
[51] V. G. Macefield, C. James, and L. A. Henderson, "Identification of sites of sympathetic outflow at rest and during emotional arousal: concurrent recordings of sympathetic nerve activity and fmri of the brain," *International Journal of Psychophysiology*, vol. 89, no. 3, pp. 451–459, 2013.
[52] J. Hayano, S. Mukai, M. Sakakibara, A. Okada, K. Takata, and T. Fujinami, "Effects of respiratory interval on vagal modulation of heart rate," *American Journal of Physiology - Heart and Circulatory Physiology*, vol. 267, no. 1, pp. H33–H40, 1994.
[53] A. P. Dutoit, E. C. Hart, N. Charkoudian, B. G. Wallin, T. B. Curry, and M. J. Joyner, "Cardiac baroreflex sensitivity is not correlated to sympathetic baroreflex sensitivity within healthy, young humans," *Hypertension*, vol. 56, no. 6, pp. 1118–1123, 2010.
[54] C. Huang, R. Gevirtz, J. Onton, and J. R. Criado, "Investigation of vagal afferent functioning using the heartbeat event related potential," *International Journal of Psychophysiology*, vol. 131, pp. 113–123, 2018.
[55] W. W. Nichols and D. G. Edwards, "Arterial elastance and wave reflection augmentation of systolic blood pressure: deleterious effects and implications for therapy," *Journal of Cardiovascular Pharmacology and Therapeutics*, vol. 6, no. 1, pp. 5–21, 2001.
[56] M. O'Rourke, "Arterial stiffness, systolic blood pressure, and logical treatment of arterial hypertension," *Hypertension*, vol. 15, no. 4, pp. 339–347, 1990.
[57] C. Marsh, *Introduction to Continuous Entropy*, Princeton University, 2013.
[58] A. Golan and J. Harte, "Information theory: a foundation for complexity science," *Proceedings of the National Academy of Sciences of the United States of America*, vol. 119, no. 33, Article ID e2119089119, 2022.
[59] J. N. Darroch and D. Ratcliff, "Generalized iterative scaling for log-linear models," *The Annals of Mathematical Statistics*, vol. 43, no. 5, pp. 1470–1480, 1972.
[60] M. E. Newman and G. T. Barkema, *Monte Carlo Methods in Statistical Physics*, Clarendon Press, 1999.
[61] E. Schneidman, S. Still, M. J. Berry, and W. Bialek, "Network information and connected correlations," *Physical Review Letters*, vol. 91, no. 23, Article ID 238701, 2003.
[62] D. Laude, J.-L. Elghozi, A. Girard et al., "Comparison of various techniques used to estimate spontaneous baroreflex sensitivity (the eurobavar study)," *American Journal of Physiology - Regulatory, Integrative and Comparative Physiology*, vol. 286, no. 1, pp. R226–R231, 2004.
[63] A. Schumann and K.-J. Bär, "Autonomic aging–a dataset to quantify changes of cardiovascular autonomic function during healthy aging," *Scientific Data*, vol. 9, pp. 95–5, 2022.
[64] A. L. Goldberger, L. A. Amaral, L. Glass et al., "Physiobank, physiotoolkit, and physionet: components of a new research resource for complex physiologic signals," *Circulation*, vol. 101, no. 23, pp. e215–E220, 2000.
[65] J. A. Behar, A. A. Rosenberg, I. Weiser-Bitoun et al., "Physiozoo: a novel open access platform for heart rate variability analysis of mammalian electrocardiographic data," *Frontiers in Physiology*, vol. 9, p. 1390, 2018.
[66] R. M. Hamilton, P. S. Mckechnie, and P. W. Macfarlane, "Can cardiac vagal tone be estimated from the 10-second ecg?" *International Journal of Cardiology*, vol. 95, no. 1, pp. 109–115, 2004.
[67] G. Parati, M. Di Rienzo, and G. Mancia, "How to measure baroreflex sensitivity: from the cardiovascular laboratory to daily life," *Journal of Hypertension*, vol. 18, no. 1, pp. 7–19, 2000.
[68] Y. Benjamini and Y. Hochberg, "Controlling the false discovery rate: a practical and powerful approach to multiple testing," *Journal of the Royal Statistical Society - Series B: Statistical Methodology*, vol. 57, no. 1, pp. 289–300, 1995.
[69] Y. Yaniv, A. E. Lyashkov, and E. G. Lakatta, "Impaired signaling intrinsic to sinoatrial node pacemaker cells affects heart rate variability during cardiac disease," *Journal of Clinical Trials*, vol. 4, no. 1, 2014.
[70] P. A. Lanfranchi and V. K. Somers, "Arterial baroreflex function and cardiovascular variability: interactions and implications," *American Journal of Physiology - Regulatory, Integrative and Comparative Physiology*, vol. 283, no. 4, pp. R815–R826, 2002.
[71] H. A. Abhishekh, P. Nisarga, R. Kisan et al., "Influence of age and gender on autonomic regulation of heart," *Journal of Clinical Monitoring and Computing*, vol. 27, no. 3, pp. 259–264, 2013.
[72] H. Nahman-Averbuch, L. Dayan, E. Sprecher et al., "Sex differences in the relationships between parasympathetic activity and pain modulation," *Physiology and Behavior*, vol. 154, pp. 40–48, 2016.
[73] R. E. De Meersman and P. K. Stein, "Vagal modulation and aging," *Biological Psychology*, vol. 74, no. 2, pp. 165–173, 2007.
[74] D. M. Kaye and M. D. Esler, "Autonomic control of the aging heart," *NeuroMolecular Medicine*, vol. 10, no. 3, pp. 179–186, 2008.
[75] K. D. Monahan, "Effect of aging on baroreflex function in humans," *American Journal of Physiology - Regulatory, Integrative and Comparative Physiology*, vol. 293, no. 1, pp. R3–R12, 2007.
[76] I. Skrapari, N. Tentolouris, and N. Katsilambros, "Baroreflex function: determinants in healthy subjects and disturbances in diabetes, obesity and metabolic syndrome," *Current Diabetes Reviews*, vol. 2, no. 3, pp. 329–338, 2006.
[77] L. Norcliffe-Kaufmann, "The vagus and glossopharyngeal nerves in two autonomic disorders," *Journal of Clinical Neurophysiology*, vol. 36, no. 6, pp. 443–451, 2019.
[78] L. Norcliffe-Kaufmann, F. Axelrod, and H. Kaufmann, "Afferent baroreflex failure in familial dysautonomia," *Neurology*, vol. 75, no. 21, pp. 1904–1911, 2010.
[79] K. Iio, S. Sakurai, T. Kato et al., "Endomyocardial biopsy in a patient with hemorrhagic pheochromocytoma presenting as inverted takotsubo cardiomyopathy," *Heart and Vessels*, vol. 28, no. 2, pp. 255–263, 2013.
[80] S. A. Klassen, D. Chirico, K. S. Dempster, J. K. Shoemaker, and D. D. O'Leary, "Role of aortic arch vascular mechanics in cardiovagal baroreflex sensitivity," *American Journal of Physiology - Regulatory, Integrative and Comparative Physiology*, vol. 311, no. 1, pp. R24–R32, 2016.
[81] I. Bonyhay, G. Jokkel, and M. Kollai, "Relation between baroreflex sensitivity and carotid artery elasticity in healthy humans," *American Journal of Physiology - Heart and Circulatory Physiology*, vol. 271, no. 3, pp. H1139–H1144, 1996.
[82] M. T. La Rovere and G. D. Pinna, "Beneficial effects of physical activity on baroreflex control in the elderly," *Annals of Noninvasive Electrocardiology*, vol. 19, no. 4, pp. 303–310, 2014.

[83] T. Laitinen, J. Hartikainen, E. Vanninen, L. Niskanen, G. Geelen, and E. Länsimies, "Age and gender dependency of baroreflex sensitivity in healthy subjects," *Journal of Applied Physiology*, vol. 84, no. 2, pp. 576–583, 1998.

[84] I. J. Scheffers, A. A. Kroon, J. Schmidli et al., "Novel baroreflex activation therapy in resistant hypertension: results of a european multi-center feasibility study," *Journal of the American College of Cardiology*, vol. 56, no. 15, pp. 1254–1258, 2010.

[85] M. J. Shen and D. P. Zipes, "Interventional and device-based autonomic modulation in heart failure," *Heart Failure Clinics*, vol. 11, no. 2, pp. 337–348, 2015.

[86] E. C. Sandset, X. Wang, C. Carcel et al., "Sex differences in treatment, radiological features and outcome after intracerebral haemorrhage: pooled analysis of intensive blood pressure reduction in acute cerebral haemorrhage trials 1 and 2," *European stroke journal*, vol. 5, no. 4, pp. 345–350, 2020.

[87] M. Sykora, J. Diedler, P. Turcani, W. Hacke, and T. Steiner, "Baroreflex: a new therapeutic target in human stroke?" *Stroke*, vol. 40, no. 12, pp. e678–e682, 2009.

[88] Q. Fu and S. Ogoh, "Sex differences in baroreflex function in health and disease," *The Journal of Physiological Sciences*, vol. 69, no. 6, pp. 851–859, 2019.

[89] R. M. Nethononda, A. J. Lewandowski, R. Stewart et al., "Gender specific patterns of age-related decline in aortic stiffness: a cardiovascular magnetic resonance study including normal ranges," *Journal of Cardiovascular Magnetic Resonance*, vol. 17, no. 1, p. 20, 2015.

[90] M. E. Mendelsohn and R. H. Karas, "Molecular and cellular basis of cardiovascular gender differences," *Science*, vol. 308, no. 5728, pp. 1583–1587, 2005.

[91] R. E. Maser and M. J. Lenhard, "An overview of the effect of weight loss on cardiovascular autonomic function," *Current Diabetes Reviews*, vol. 3, pp. 204–211, 2007.

[92] S. Grewal and V. Gupta, "Effect of obesity on autonomic nervous system," *International Journal of Current Biological and Medical Science*, vol. 1, pp. 15–18, 2011.

[93] B.-Y. Li, X.-Y. Li, X. Lu, R. Kang, Z.-X. Tian, and F. Ling, *Utilizing Entropy to Systematically Quantify the Resting-Condition Baroreflex Regulation Function*, 2023.